\documentclass[aps, 
superscriptaddress, reprint, twocolumn,
amsmath, amssymb
]{revtex4-2}

\usepackage{ragged2e}
\usepackage{xparse}
\usepackage{mathtools}
\usepackage{bbold}
\usepackage{tikz}
\usetikzlibrary{arrows,arrows.meta,patterns,decorations.pathreplacing,external}
\usepackage{qcircuit}
\usepackage{physics}
\usepackage{romannum}
\usepackage{blkarray}
\usepackage[caption=false]{subfig}
\usepackage[hidelinks]{hyperref}
\hypersetup{colorlinks=true,
            urlcolor=blue,
            linkcolor=black,
            citecolor=blue}

\usepackage[shortlabels]{enumitem} 

\newcommand{\toremove}[1]{}
\newcommand{\z}{\mathbf{z}}
\newcommand{\T}{\mathbb{T}}
\newcommand{\R}{\mathbb{R}}
\newcommand{\N}{\mathbb{N}}
\newcommand{\C}{\mathbb{C}}
\newcommand{\Z}{\mathbb{Z}}

\begin{document}
\pagenumbering{arabic}

\title{Chaotic Roots of the Modular Multiplication Dynamical System in Shor's Algorithm}
\author{Abu Musa Patoary}
\affiliation{Joint Quantum Institute and Department of Physics, University of Maryland, College Park, MD 20742, USA}
\author{Amit Vikram}
\affiliation{Joint Quantum Institute and Department of Physics, University of Maryland, College Park, MD 20742, USA}
\author{Laura Shou}
\affiliation{School of Mathematics, University of Minnesota, Minneapolis, MN 55455, USA}
\author{Victor Galitski}
\affiliation{Joint Quantum Institute and Department of Physics, University of Maryland, College Park, MD 20742, USA}
\affiliation{Center for Computational Quantum Physics, The Flatiron Institute, New York, NY 10010, United States}

\begin{abstract} Shor's factoring algorithm, believed to provide an exponential speedup over classical computation, relies on finding the period of an exactly periodic quantum modular multiplication operator. This exact periodicity is the hallmark of an integrable system, which is paradoxical from the viewpoint of quantum chaos, given that the classical limit of the modular multiplication operator is a highly chaotic system that occupies the ``maximally random'' Bernoulli level of the classical ergodic hierarchy. In this work, we approach this apparent paradox from a quantum dynamical systems viewpoint, and consider whether signatures of ergodicity and chaos may indeed be encoded in such an ``integrable'' quantization of a chaotic system.
We show that Shor's modular multiplication operator, in specific cases, can be written as a superposition of quantized $A$-baker's maps exhibiting more typical signatures of quantum chaos and ergodicity. This work suggests that the integrability of Shor's modular multiplication operator may stem from the interference of other ``chaotic'' quantizations of the same family of maps, and paves the way for deeper studies on the interplay of integrability, ergodicity and chaos in and via quantum algorithms.
\end{abstract}
\date{\today}

\maketitle

\textit{Introduction} --- {Shor's algorithm~\cite{shor1994algorithms, shor1999polynomial} to factorize an integer $N$ is a cornerstone of quantum computation~\cite{NielsenChuang}, being exponentially faster than all known classical factorization algorithms and capable of breaking RSA encryption~\cite{rivest1978method}, a widely used scheme for secure data transmission.} 
{Interestingly, the success of this algorithm hinges on a foundational tension with the very notion of ``quantum chaos''~\cite{Haake}: the quantum modular multiplication operator at the core of Shor's algorithm~\cite{shor1994algorithms, shor1999polynomial, NielsenChuang} belongs to a class of quantum systems that have a strongly chaotic classical limit, but  violate several expected signatures of ``chaos'' on quantization.}

The quantization of classically ergodic, chaotic systems~\cite{Haake, Ott} is typically associated with spectral signatures such as non-degenerate energy levels
and spectral rigidity, as supported by general theoretical arguments and numerical studies of several systems~\cite{Haake, DKchaos, CGV, BerryStadium, BGS, HOdA, BerrySR, argaman, HaakePO, HaakePO2, dynamicalqergodicity}.
Paradoxically, as indicated above, some quantizations~\cite{shor1994algorithms, hannay1980quantization} of certain canonical textbook examples of classically ergodic and chaotic systems~\cite{Ott} -- such as modular multiplication $f_A(x) = (A x \mod N)$ on a 1D interval $x\in [0,N)$ (quantized in Shor's algorithm) and Arnold's cat map on the torus --- are \textit{exactly periodic} at long times with highly degenerate and orderly spectra~\cite{shor1994algorithms, keating1991cat}, which is at odds with ergodic quantum dynamics~\cite{dynamicalqergodicity}. This is because quantization itself is not a uniquely defined procedure, and several distinct quantum systems can have the same classical limit (e.g., Refs.~\cite{shor1994algorithms, QuantumGraphs1D, balazs1989quantized, saraceno1990classical} consider entirely different quantizations of $f_2(x)$ with different spectral properties). In particular, the (conventionally) ``standard'' quantization of each of the above maps~\cite{shor1994algorithms, hannay1980quantization} captures the dynamics of only a \textit{measure-zero} subset of periodic orbits with a common (not necessarily fundamental) period~\cite{dysoncat, Ott}. This gives the appearance of early-time ``chaos'' on quantization via the exponential divergence of typical nearby orbits, but completely misses out on the full ergodic and chaotic classical dynamics at late times in the bulk of the phase space.

On the one hand, the example of Shor's algorithm demonstrates a possibly generic need to eliminate ``quantum chaos'' in the quantization of even a classically chaotic map, for its successful utilization in certain quantum algorithms. On the other, it also suggests the more fundamental question of whether appropriate manifestations of quantum ergodicity and chaos can be hidden in some way even in such non-ergodic quantizations.
Interestingly, a partial resolution to this question was noted in Refs.~\cite{lakshminarayan2005shuffling, lakshminarayan2007modular}, where it was shown~\cite{lakshminarayan2007modular} that the unitary operator implementing modular multiplication by $A=2$ in Shor's algorithm can be expressed as a superposition of quantized baker's maps~\cite{balazs1987quantized, balazs1989quantized}, which may be regarded as ``chaotic'' quantizations of $f_2(x)$ [more precisely, of baker's maps~\cite{Ott} in a 2D phase space $(x,p)$ whose action on the position coordinate $x \in [0,1)$ is identical to $f_2(Nx)/N$] that by and large exhibit the expected signatures of quantum chaos and ergodicity~\cite{balazs1989quantized, saraceno1990classical}.

In this work, we show that this ``embedding'' of a superposition of ``quantum chaotic'' maps in the periodic modular multiplication map generalizes to an arbitrary multiplier $A$. The appropriate chaotic maps are direct generalizations of the 2D $A$-baker's maps (extensions of baker's maps whose 1D projection is $f_A(Nx)/N$), and these maps can be quantized in terms of certain combinations of discrete Fourier transforms~\cite{balazs1989quantized}, when either $N+1$ or $N-1$ is a multiple of $A$. This establishes a rigorous correspondence between specific ``ergodic'' and ``non-ergodic'' quantizations of $f_A(x)$ for arbitrary $A$, which may serve as a simple model for understanding the interplay of integrability, ergodicity and chaos in different quantizations of the same system.
We will now present the details of this correspondence, and subsequently discuss both its potential implications for studying the embedding of quantum chaos in Shor's algorithm, and in extending the quantization of $A$-baker's maps to arbitrary $N$.

\textit{Results} --- In the remainder of this paper, we will refer to the exactly periodic quantum modular multiplication used in Shor's algorithm simply as modular multiplication, without explicitly mentioning the qualifier ``quantum''. Our main result (Eq.~\eqref{mod mult in terms of Bernoulli}) is to show that modular multiplication, in some specific cases, can be exactly written as a superposition of ``chaotic'' quantum \textit{$A$-baker's maps}. 
Modular multiplication $U_A$, given two co-primes $N$ and $A$ $(A < N)$, is defined in an $N$ dimensional Hilbert space according to the equation:
\begin{equation} \label{definition: Mod Multiplication}
U_A\ket{m}=\ket{mA \pmod{N}},\ \text{for \hspace{.1 cm}} m \in \lbrace 0,1 \dots N-1\rbrace,
\end{equation}
where $\{\ket{0}, \ket{1} \dots \ket{N-1}\}$ forms an orthonormal basis. Since $N$ and $A$ are co-primes, modular multiplication is an exactly periodic unitary operator which permutes the basis states according to Eq. \eqref{definition: Mod Multiplication}.

\begin{figure}[htb]
    \centering
    \includegraphics[width=.80\columnwidth]{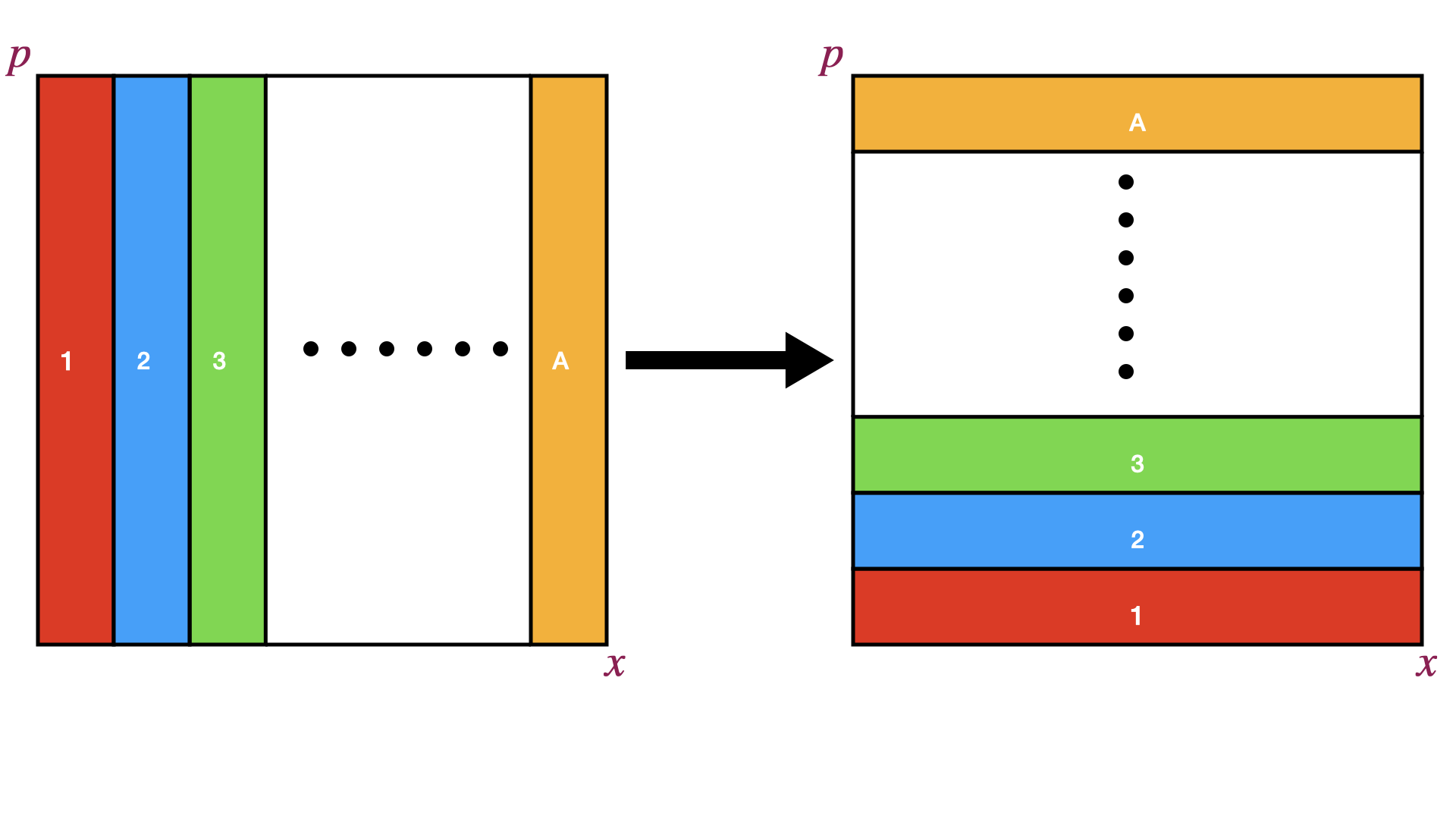}
    \caption{The A-baker's map transforms the unit square in the left to the one in the right. The unit square is divided into $A$ rectangles which are marked by numbers from $1$ to $A$. Each rectangle is stretched along $x$ axis and contracted along $p$ axis before being stacked on top of each other.}
    \label{fig:1}
\end{figure}

The classical $A$-baker's map, a generalization of baker's map, is a canonical example of an ergodic and chaotic system~
\cite{AN}. It maps a unit square to itself by stretching in one direction and compressing in the perpendicular direction in such a way that the total area of the square is preserved, and rearranging the stretched square to fit inside the unit square. The map is defined by the following transformation:
\begin{align} \label{definition: Bernoulli Map q}
    x \rightarrow x^{'} = Ax - \lfloor Ax\rfloor, \\ \label{definition: Bernoulli Map p}
    p \rightarrow p^{'} = \frac{p+\lfloor Ax \rfloor}{A}.
\end{align}
Here $(x,p) \in [0,1)\cross[0,1)$ are phase space coordinates and $\lfloor x \rfloor$ denotes the integer part of $x$. The map is illustrated in Fig. \ref{fig:1}. The unit square is divided into $A$ rectangles of equal area as shown in the left side of Fig. \ref{fig:1}. Each of the rectangles is stretched by a factor of $A$ along the horizontal direction and by $1/A$ along the vertical direction. Then the rectangles are stacked on top of each other as shown on the right side of Fig. \ref{fig:1}.  
When $A=2$ the $A$-baker's map reduces to the standard baker's map~\cite{renyi1957representations}. 

As with any classical dynamical system, there is no unique way to quantize the $A$-baker's map. By convention, the basic requirements are that the quantum map is unitary and it reduces to the classical map in the semi-classical limit. Several quantization procedures satisfying these requirements have been developed for $A$-baker's map \cite{balazs1989quantized,saraceno1990classical,saraceno1994towards,rubin1998canonical,ermann2006generalized,AN}. In this paper, we will use the quantization procedure developed by Balazs and Voros (BV)\cite{balazs1989quantized}. In BV quantization, one replaces the phase space of the classical map with a $D$ dimensional Hilbert space. Then one can consider discrete position $(\ket{x_n})$ and momentum $(\ket{p_n})$ bases with the boundary conditions $\ket{x_{n+D}} =e^{-2\pi i \beta} \ket{x_n}$ and $\ket{p_{n+D}} =e^{2\pi i \alpha} \ket{p_n}$. Finally one constructs a $D \cross D$ unitary matrix which transforms the states in ways analogous to the classical transformation in Eq. \eqref{definition: Bernoulli Map q} and \eqref{definition: Bernoulli Map p}. 
This unitary matrix is the quantum $A$-baker's map. The Hilbert space dimension $D$ plays the role of $\hbar^{-1}$~\cite{balazs1989quantized}. Therefore the semi-classical limit of this map is obtained by taking $D \to \infty$ which is equivalent to $\hbar \to 0$. We explain the associated procedures in detail in Sections~\ref{sec:detail} and \ref{sec:limit} of the supplemental material. The quantum $A$-baker's map, obtained using the BV procedure, is given by the following unitary matrix $B_A^{(0)}$: 
\begin{equation} \label{quantized Bernoulli map}
    B_A^{(0)} = (F_D^{\alpha, \beta})^{-1} \begin{bmatrix}
        F_{\frac{D}{A}}^{\alpha, \beta} & 0 & 0 & \dots & 0 \\
        0 & F_{\frac{D}{A}}^{\alpha, \beta} & 0 & \dots & 0 \\
        0 & 0 & F_{\frac{D}{A}}^{\alpha, \beta} & \dots & 0 \\
        \vdots & \vdots & \vdots & \ddots & \vdots \\
        0 & 0 & 0 & \dots & F_{\frac{D}{A}}^{\alpha, \beta}
    \end{bmatrix}.
\end{equation}
Here $\alpha, \beta$ are phases introduced in the boundary condition of position and momentum basis and $F_D^{\alpha, \beta}$ are generalized discrete Fourier transform (DFT) matrix whose elements are 
\begin{equation} \label{DFT elements}
 [F_D^{\alpha, \beta}]_{nm} = \frac{1}{\sqrt{D}}e^{-2\pi i (n+\alpha) (m+\beta)/D}.  
\end{equation} 
For periodic boundary conditions ($\alpha=\beta=0$), $F_D^{\alpha, \beta}$ reduces to the standard DFT matrix. For notational convenience we will denote the standard DFT matrix simply by $F_D$. Note that $D/A$ in Eq. \eqref{quantized Bernoulli map} should be an integer which requires $D$ to be a multiple of $A$. 

One can modify the $A$-baker's map so that the rectangles in the right side of Fig. \ref{fig:1} are stacked differently, and then use the same BV procedure to quantize this modified map. For example, if all the rectangles in the right side of Fig. \ref{fig:1} are cyclically permuted by one step so that now $1$ goes to $A$, $2$ goes to $1$ \dots $3$ goes to $2$, we get a different quantized map $B_A^{(1)}$. 
Similarly one can construct other quantum maps $B_A^{(2)}, B_A^{(3)} \dots B_A^{(A-1)}$ by cyclically permuting the horizontal rectangles by $2$, $3$,$\dots$,$A-1$ steps respectively. The general equation for the $A$-baker's map $B_A^{(k)}$ is
\begin{equation} \label{permuted quantized Bernoulli}
B_A^{(k)} =  [F_D^{\alpha, \beta}]^{-1}
\begin{blockarray}{cccccc}
  0 &\dots & k-1 & k & \dots & A-1 \\
\begin{block}{[cccccc]}
   0  & \dots & 0 & F_{\frac{D}{A}}^{\alpha,\beta} & \dots  & 0   \\
   \vdots & \ddots & \vdots & \vdots & \ddots & \vdots \\  
   0  & \dots  & 0 & 0 & \dots & F_{\frac{D}{A}}^{\alpha,\beta} \\
   F_{\frac{D}{A}}^{\alpha,\beta}  & \dots  & 0 & 0 & \dots & 0 \\
   \vdots & \ddots &  \vdots & \vdots & \ddots & \vdots  \\ 
   0  & \dots  & F_{\frac{D}{A}}^{\alpha,\beta} & 0 & \dots & 0  \\
\end{block}
\end{blockarray}\hspace{.1 cm},
\end{equation}
where $k\in\{0,1,.....A-1\}$. When $k=0$ it reduces to Eq. \eqref{quantized Bernoulli map}.

With this background we can state our main result. 
Given two co-prime integers $N$ and $A$ such that $N = Aq \pm 1$ where $q$ is an integer, the corresponding modular multiplication operator $U_A^{\pm}$ can written to act instead on $D=Aq$ states (rather than on $N=Aq\pm1$ as usual). This operator is obtained in the case $N=Aq+1$ by removing the state $|0\rangle$ which is always a fixed point, $U_A^+|0\rangle=|0\rangle$, and in the case $N=Aq-1$ by adding a $D$th state $|D-1\rangle$, with the extension $U_A^-|D-1\rangle = |D-1\rangle$ which agrees with the original map defined in Eq.~\eqref{definition: Mod Multiplication}. The modular multiplication map dynamics are thus preserved, but the advantage is the number of states is now a multiple of $A$.
With these conventions, the $Aq\times Aq$ representation of the modular multiplication operator $U_A^\pm$ can be expressed as
\begin{equation} \label{mod mult in terms of DFT}
    U_A^{\pm} = F_{Aq}^{-1} \begin{bmatrix} F_q & F_q^{0,\mp \frac{1}{A}} & F_q^{0,\mp \frac{2}{A}} & \dots & F_q^{0, \mp \frac{A-1}{A}}\\ F_q &  F_q^{0,\mp \frac{1}{A}} &  F_q^{0,\mp \frac{2}{A}} & \dots &  F_q^{0,\mp \frac{A-1}{A}} \\ F_q &  F_q^{0,\mp \frac{1}{A}} &  F_q^{0,\mp \frac{2}{A}} & \dots &  F_q^{0,\mp \frac{A-1}{A}} \\ \vdots & \vdots & \vdots & \vdots & \vdots \\ F_q &  F_q^{0,\mp \frac{1}{A}} &  F_q^{0,\mp \frac{2}{A}} & \dots &  F_q^{0,\mp \frac{A-1}{A}}\end{bmatrix}   \odot \Tilde{F}_{Aq}^{\mp}. 
\end{equation}
Here $\odot$ denotes Hadamard or entry-wise product of two matrices defined as $(A\odot B)_{ij} = A_{ij}B_{ij}$ and $\Tilde{F}_{Aq} = F_A^{\mp}\otimes J_q$ with $J_q$ being a $q \cross q$ matrix with all elements equal to $1$, $F_A^{+}$ the $A \cross A$ DFT matrix with no phase, $F_A^{-}$ the conjugate of $F_A^{+}$ and $\otimes$ denoting tensor product.
In Section~\ref{sec:analytic} of the supplemental material we analytically derive Eq. \eqref{mod mult in terms of DFT}.

To extract the quantized $A$-baker's maps embedded in Eq. \eqref{mod mult in terms of DFT}, we rewrite it as:
\begin{equation} \label{mod mult in terms of Bernoulli}
    U_A^{\pm} = \frac{1}{\sqrt{A}} \sum_{k=0}^{A-1} \Tilde{B}_{A}^{\pm (k)},
\end{equation}
where $\Tilde{B}_{A}^{\pm (k)}$ are quantized maps similar to the ones obtained in Eq. \eqref{permuted quantized Bernoulli} but with some differences. Let us write the explicit form of $\Tilde{B}_{A}^{\pm(k)}$ to identify the differences:
\begin{equation} \label{Bernoulli tilde}
\Tilde{B}_A^{\pm (k)} \!\!=  F_{Aq}^{-1}
\begin{blockarray}{cccccc}
  0 &\dots & k-1 & k & \dots & A-1 \\
\begin{block}{[cccccc]}
    0  & \dots & 0 & F_{q}^{0,\mp \frac{k}{A}} & \dots  & 0   \\
   \vdots & \ddots & \vdots & \vdots & \ddots & \vdots  \\  
   0  & \dots  & 0 & 0 & \dots & F_{q}^{0,\mp \frac{A-1}{A}}   \\
   F_ {q} & \dots  & 0 & 0 & \dots & 0   \\
   \vdots & \ddots &  \vdots & \vdots & \ddots & \vdots  \\ 
   0  & \dots  & F_{q}^{0,\mp \frac{k-1}{A}} & 0 & \dots & 0   \\
\end{block}
\end{blockarray} \odot \Tilde{F}_{Aq}^{\mp}. 
\end{equation}

\begin{figure}[htb]
    \centering
    \includegraphics[width=.80\columnwidth]{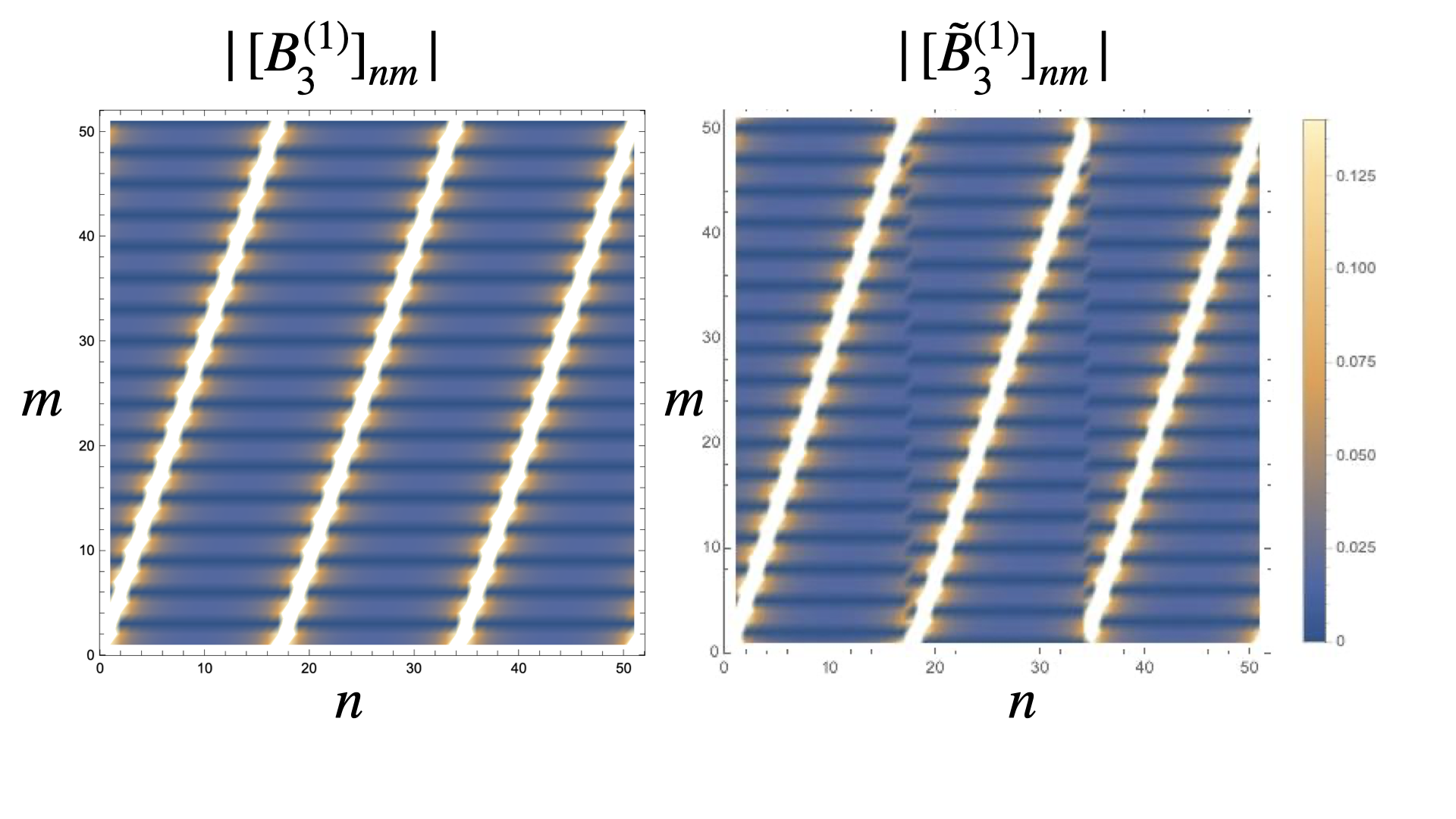}
    \caption{The density plot of the absolute value of the matrix elements of $B_3^{(1)}$ (left) and $\Tilde{B}_3^{(1)}$(right) when $N=52$. They are mostly identical except minor difference in the cuts between the 3 segments. Both of them resemble the classical A-baker's map for $A=3$. This is necessary but not sufficient to claim that $\Tilde{B}_3^{(1)}$ reduces to 3-baker's map in semi-classical limit. 
    }
    \label{fig:2}
\end{figure}

Comparing Eq. \eqref{Bernoulli tilde} with Eq. \eqref{permuted quantized Bernoulli} we notice two main differences: (\romannum{1}) In $\Tilde{B}_{A}^{\pm (k)}$, the inverse of $F_{Aq}$ has no phase and the phases of each $q \cross q$ DFT matrices are different but in $B_A^{(k)}$ all of them have same phase, (\romannum{2}) The Hadamard product in Eq. \eqref{Bernoulli tilde} is absent in Eq. \eqref{permuted quantized Bernoulli}. 
Despite these differences, the evolution of states under the operators $\Tilde{B}_{A}^{\pm (k)}$ is very similar to that of $B_A^{(k)}$ in the semi-classical limit, and we will call $\Tilde{B}_{A}^{\pm (k)}$ \emph{quantum $A$-baker's maps}. We note that if we plot the matrix elements of $\Tilde{B}_A^{\pm(k)}$ they resemble classical Bernoulli maps in the same way as the $B_A^{(k)}$ do (Fig.~\ref{fig:2}), although as we explain shortly this is not a sufficient criterion.

\begin{figure}[htb]
\subfloat[\justifying Phase space plots (Husimi functions) of the quantum evolution of a Gaussian state $|\Psi\rangle$ by $\tilde{B}_3^{+(1)}$, for $D=150$. The state $|\Psi\rangle$ is localized at $(x,p)=(\frac{1}{2},\frac{1}{2})$,  and its time evolved state $\tilde{B}_3^{+(1)}|\Psi\rangle$ is localized near the classical trajectory point $(3x,\frac{p+(1-1)}{3})=(\frac{1}{2},\frac{1}{6})$. The colored rectangles are overlayed for easier comparison to the classical map in (b).
]{\includegraphics[width=.42\textwidth]{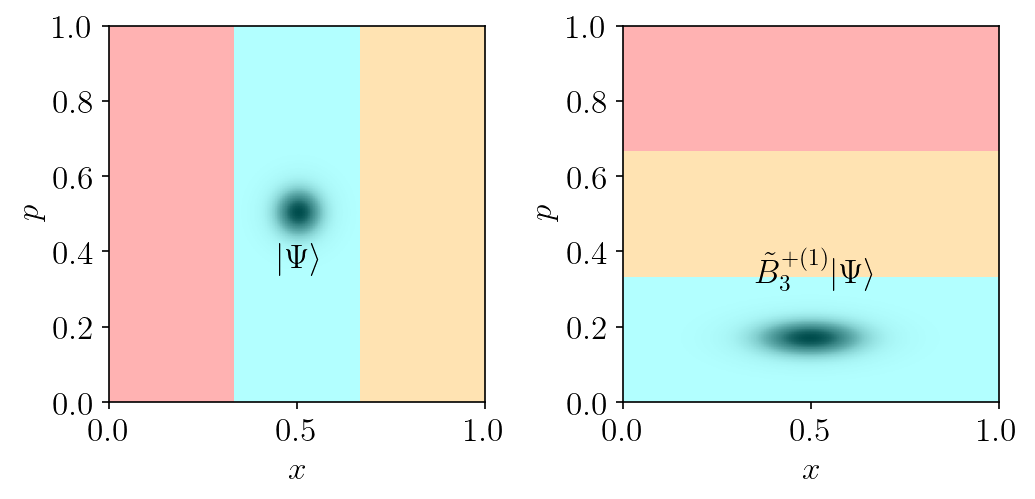}\label{subfig:q}
}\\
\subfloat[\justifying Classical evolution of the unit square by the $3$-baker map with cyclic shift $k=1$,  which sends $(x,p)\mapsto (3x-\lfloor 3x\rfloor,\frac{p+(\lfloor Ax\rfloor-1)\;\mathrm{mod}\,A}{A})$. The total classical transformation is shown in the top row, while the intermediate components of stretching, cutting and stacking, and cyclic permutation are shown through the two stages in the bottom row.
The action on a disk is also shown for comparison to part (a). 
]{\begin{tikzpicture}[scale=0.28]
\def\alpha{30} 
\def\radius{.04cm} 
\def\xcoord{.5cm} 
\def\ycoord{.5cm} 
\def\beta{100} 
\def\col{black} 
\def\drawcol{black} 
\def\globalscale{.28}
\def\first{\begin{tikzpicture}[scale=\globalscale]
\node [below left] at (0,0) {$0$};
\node [below] at (6,0) {$1$};
\node [left] at (0,6) {$1$};
\node [below] at (2,0) {$\frac{1}{3}$};
\draw (2,0)--(2,6);
\node [below] at (4,0) {$\frac{2}{3}$};
\draw (4,0)--(4,6);
\fill[color=red!\alpha] (0,0)--(2,0)--(2,6)--(0,6)--cycle;
\fill[color=cyan!\alpha] (2,0)--(4,0)--(4,6)--(2,6)--cycle;
\fill[color=orange!\alpha] (4,0)--(6,0)--(6,6)--(4,6)--cycle;
\draw (0,0)--(6,0)--(6,6)--(0,6)--cycle;
\node [fill, draw=\drawcol,circle, minimum size = \radius*6, inner color = black, outer color = \col!\beta, inner sep = 0pt] at (\xcoord*6,\ycoord*6) {};
\end{tikzpicture}
}
\def\second{\begin{tikzpicture}[scale=\globalscale]
\def\xtwo{9cm};
\node [below left] at (0,0) {$0$};
\node [below] at (6,0) {$1$};
\node [below] at (12,0) {$2$};
\node [below] at (18,0) {$3$};
\node [left] at (0,2) {$\frac{1}{3}$};
\draw (6,0)--(6,2);
\draw (12,0)--(12,2);
\fill[color=red!\alpha] (0,0)--(6,0)--(6,2)--(0,2)--cycle;
\fill[color=cyan!\alpha] (6,0)--(12,0)--(12,2)--(6,2)--cycle;
\fill[color=orange!\alpha] (12,0)--(18,0)--(18,2)--(12,2)--cycle;
\draw (0,0)--(18,0)--(18,2)--(0,2)--cycle;
\node [fill, draw=\drawcol,circle, minimum size=\radius*6,yscale=1/3,xscale=3,inner color = black, outer color = \col!\beta,inner sep = 0pt] at (3*\xcoord*6,\ycoord*6/3) {};
\end{tikzpicture}
}
\def\third{\begin{tikzpicture}[scale=\globalscale]
\node [below left] at (0,0) {$0$};
\node [below] at (6,0) {$1$};
\node [left] at (0,6) {$1$};
\draw (0,2)--(6,2);
\draw (0,4)--(6,4);
\fill[color=red!\alpha] (0,0)--(6,0)--(6,2)--(0,2)--cycle;
\fill[color=cyan!\alpha,yshift=2cm] (0,0)--(6,0)--(6,2)--(0,2)--cycle;
\fill[color=orange!\alpha,yshift=4cm] (0,0)--(6,0)--(6,2)--(0,2)--cycle;
\draw (0,0)--(6,0)--(6,6)--(0,6)--cycle;
\node [fill, draw=\drawcol,circle, minimum size=\radius*6,yscale=1./3,xscale=3,inner color = black, outer color = \col!\beta,inner sep = 0pt] at (3*\xcoord*6-6cm,\ycoord*6/3+2cm) {};
\end{tikzpicture}
}
\def\fourth{\begin{tikzpicture}[scale=\globalscale]
\def\xthree{10cm};
\node [below left] at (0,0) {$0$};
\node [below] at (6,0) {$1$};
\node [left] at (0,6) {$1$};
\fill[color=cyan!\alpha] (0,0)--(6,0)--(6,2)--(0,2)--cycle;
\fill[color=orange!\alpha,yshift=2cm] (0,0)--(6,0)--(6,2)--(0,2)--cycle;
\fill[color=red!\alpha,yshift=4cm] (0,0)--(6,0)--(6,2)--(0,2)--cycle;
\draw (0,0)--(6,0)--(6,6)--(0,6)--cycle;
\node [fill, draw=\drawcol,circle, minimum size=\radius*6,yscale=1./3,xscale=3,inner color = black, outer color = \col!\beta,inner sep = 0pt] at (3*\xcoord*6-6cm,\ycoord*6/3) {};
\end{tikzpicture}
}
\node at (2,0) {\first};
\node at (6,-10.5) {\second};
\node at (21,-9) {\third};
\node at (21,0) {\fourth};
\draw[->] (1,-5)--(1,-8);
\draw[->] (16,-10.5)--(18,-10.5);
\draw[->] (21,-5)--(21,-3);
\node[right] at (21.2,-4.5) {\footnotesize{$k=1$}};
\draw [line width=1pt, double distance=.5pt, arrows={-Latex[length=0pt 3 0]},xshift=.3cm] (7,0)--(15,0);
\node[above] at (11,0.2) {\footnotesize{$3$-baker map}};
\node[below] at (11,-.2) {\footnotesize{w/ shift $k=1$}};
\end{tikzpicture}
}\label{subfig:c}
\caption{Comparison of the quantum evolution under $\tilde{B}_3^{+(1)}$, and the classical evolution by the $3$-baker map with cyclic shift $k=1$.
The time-evolved quantum state $\tilde{B}_3^{+(1)}|\Psi\rangle$ shown in (a) follows the classical action shown in (b), and also displays the stretching in $x$-direction and shrinking in $p$-direction characteristic of the classical $A$-baker map.
}\label{fig:3}
\end{figure}
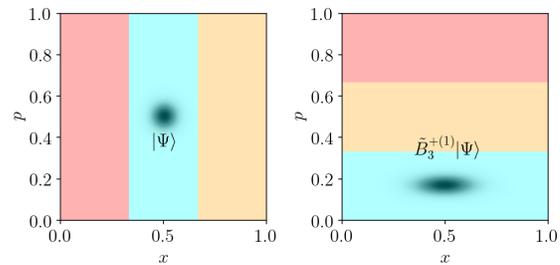
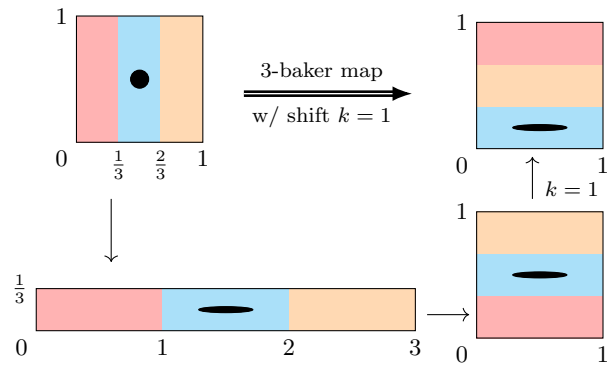

\textit{Classical limit ---} To justify calling $\tilde{B}_A^{\pm(k)}$ a quantum $A$-baker's map, in Section~\ref{sec:limit} of the supplementary material we determine the action of $\tilde{B}_A^{\pm(k)}$ on Gaussian states (``coherent states'') maximally localized at a point $(x,p)$ in phase space. We show that $\tilde{B}_A^{\pm(k)}$ sends such a state to another coherent state localized close to the classical evolution location $(Ax-\lfloor Ax\rfloor,\frac{p+(\lfloor Ax\rfloor-k)\;\mathrm{mod}A}{A})$.
We only do this for $(x,p)$ away from the discontinuities of the classical $A$-baker map, otherwise there can be diffraction effects~\cite{saraceno1994towards,DBDE,DNW}.
The action on coherent states can be interpreted in terms of the Wigner or Husimi function in phase space: A state whose Wigner/Husimi function is localized near $(x,p)$ is transformed by $\tilde{B}_A^{\pm(k)}$ to a state localized near the classical trajectory point $(Ax-\lfloor Ax\rfloor,\frac{p+(\lfloor Ax\rfloor-k)\;\mathrm{mod}A}{A})$. This correspondence between quantum and classical evolution is shown visually in Fig.~\ref{fig:3}.

Such an action on Gaussian states was used in \cite{DNW} to  prove a rigorous classical-quantum correspondence (Egorov-type theorem \cite[Theorem 12]{DNW}) for the original Balazs--Voros quantization of the baker map.  
In the case here, we will be content with just analyzing the behavior of $\tilde{B}_A^{\pm(k)}$ on Gaussian states.
Due to the varying phases in the DFT matrices in Eq.~\eqref{Bernoulli tilde}, the action on Gaussian states will be more complicated than in the original Balazs--Voros quantization. In particular, the quantum evolved state here is actually centered a distance $\mathcal{O}_A(D^{-1})$ away from the classical trajectory point, though this will still be close enough for our purposes to the Gaussian state actually centered at the classical trajectory.


As noted previously, the matrix elements of $\tilde{B}_A^{\pm(k)}$ in the position basis (and momentum basis) trace out the classical 
1D action of the position (momentum) coordinate in the $A$-baker map defined in Eqs.~(\ref{definition: Bernoulli Map q}-\ref{definition: Bernoulli Map p}). This is not sufficient to conclude that $\tilde{B}_A^{\pm(k)}$ has the correct semi-classical behavior, as the Walsh baker quantizations considered in \cite{AN} also display this behavior, but as shown in \cite{TracyScott} are not quantizations of the baker map. Instead the semi-classical limit there has to be interpreted as a ``stochastic'' or ``multi-valued'' baker map. One explanation for this is that even if the maps look classically correct in pure position or pure momentum basis, they may not entwine position and momentum together correctly. 
Thus checking the behavior in phase space, on coherent states or through Wigner/Husimi functions, is necessary to understand the semiclassical limit.

\textit{Discussion} --- We have illustrated a specific ``embedding'' of quantum chaotic behavior in the periodic orbits of a classically ergodic, chaotic map via quantum superposition, by deriving an exact correspondence between the periodic modular-multiplication-by-$A$ maps and ``chaotic'' $A$-baker's maps, for $N$ such that either of $N\pm 1$ is a multiple of $A$.  As will be shown in an upcoming work, this result generalizes to other values of $N$ albeit via a more complicated construction with two important implications.

Firstly, in any practical implementation of Shor's algorithm, it is $N$ that is given while $A$ is chosen as per convenience, in contrast to the dynamical systems approach, in which a fixed $A$ specifies the modular multiplication map while $N\to\infty$ through any subsequence of $\mathbb{N}$ gives the classical limit. This means that a general result for $N$ would be crucial if one is to understand the implications of this ``embedding of quantum chaos'' for Shor's algorithm (otherwise, choosing an $A$ that is a factor of either of $N\pm 1$ would require the output of Shor's algorithm as applied to $N\pm 1$, introducing a circular element). Conversely, achieving such a generalization may provide a fuller picture of how certain perturbations affect the dynamics of Shor's algorithm, along the lines of the analysis in Ref.~\cite{lakshminarayan2007modular}.

Secondly, from a fundamental quantum dynamical systems viewpoint, the study of quantized $A$-baker's maps has generally been restricted to Hilbert space dimensions that are multiples of $A$, due to a need to consider $A$ copies of discrete Fourier transforms in, e.g., the Balazs--Voros quantization~\cite{balazs1989quantized}. Approaching the quantization of such maps through the above connection to Shor's algorithm, on the other hand, provides a promising avenue to generalize $A$-baker's maps to a wider set of Hilbert space dimensions (or chocies of $N$) with the use of \textit{generalizations} of Fourier transforms, in particular for any coprime pair of $(A,N)$. This also opens avenues for studying the semiclassical behavior of such generalized quantum $A$-baker's maps to explore atypicalities and any $N$-dependence in their ``quantum chaos'' signatures, noting that such atypicalities were observed for $N$ a multiple of $A$ in the Balazs-Voros quantization~\cite{balazs1989quantized}. Further, $A$-baker's maps can be directly realized in quantum simulators using quantum Fourier transforms~\cite{bakingprotocol, bakingsimulationNMR}, which may allow the experimental detection of such atypical spectral signatures using recently developed measurement protocols for signatures of ``quantum chaos''~\cite{QNDSFF, pSFF}. 

{\em Acknowledgments} -- This work was primarily supported by the U.S. Department of Energy, Office of Science, Basic Energy Sciences under Award No. DE-SC0001911 (V.G. and A.V., theory of quantum chaos). A.P. (establishing a connection between quantum maps and Shor's algorithm) acknowledges support from the National Science Foundation (QLCI grant OMA-2120757). L.S. was supported by Simons Foundation grant 563916, SM.

\bibliography{references} 

\begin{thebibliography}{38}%
\makeatletter
\providecommand \@ifxundefined [1]{%
 \@ifx{#1\undefined}
}%
\providecommand \@ifnum [1]{%
 \ifnum #1\expandafter \@firstoftwo
 \else \expandafter \@secondoftwo
 \fi
}%
\providecommand \@ifx [1]{%
 \ifx #1\expandafter \@firstoftwo
 \else \expandafter \@secondoftwo
 \fi
}%
\providecommand \natexlab [1]{#1}%
\providecommand \enquote  [1]{``#1''}%
\providecommand \bibnamefont  [1]{#1}%
\providecommand \bibfnamefont [1]{#1}%
\providecommand \citenamefont [1]{#1}%
\providecommand \href@noop [0]{\@secondoftwo}%
\providecommand \href [0]{\begingroup \@sanitize@url \@href}%
\providecommand \@href[1]{\@@startlink{#1}\@@href}%
\providecommand \@@href[1]{\endgroup#1\@@endlink}%
\providecommand \@sanitize@url [0]{\catcode `\\12\catcode `\$12\catcode
  `\&12\catcode `\#12\catcode `\^12\catcode `\_12\catcode `\%12\relax}%
\providecommand \@@startlink[1]{}%
\providecommand \@@endlink[0]{}%
\providecommand \url  [0]{\begingroup\@sanitize@url \@url }%
\providecommand \@url [1]{\endgroup\@href {#1}{\urlprefix }}%
\providecommand \urlprefix  [0]{URL }%
\providecommand \Eprint [0]{\href }%
\providecommand \doibase [0]{https://doi.org/}%
\providecommand \selectlanguage [0]{\@gobble}%
\providecommand \bibinfo  [0]{\@secondoftwo}%
\providecommand \bibfield  [0]{\@secondoftwo}%
\providecommand \translation [1]{[#1]}%
\providecommand \BibitemOpen [0]{}%
\providecommand \bibitemStop [0]{}%
\providecommand \bibitemNoStop [0]{.\EOS\space}%
\providecommand \EOS [0]{\spacefactor3000\relax}%
\providecommand \BibitemShut  [1]{\csname bibitem#1\endcsname}%
\let\auto@bib@innerbib\@empty
\bibitem [{\citenamefont {Shor}(1994)}]{shor1994algorithms}%
  \BibitemOpen
  \bibfield  {author} {\bibinfo {author} {\bibfnamefont {P.~W.}\ \bibnamefont
  {Shor}},\ }\bibfield  {title} {\bibinfo {title} {Algorithms for quantum
  computation: discrete logarithms and factoring},\ }in\ \href
  {https://doi.org/10.1109/SFCS.1994.365700} {\emph {\bibinfo {booktitle}
  {Proceedings 35th annual symposium on foundations of computer science}}}\
  (\bibinfo {organization} {IEEE},\ \bibinfo {year} {1994})\ pp.\ \bibinfo
  {pages} {124--134}\BibitemShut {NoStop}%
\bibitem [{\citenamefont {Shor}(1999)}]{shor1999polynomial}%
  \BibitemOpen
  \bibfield  {author} {\bibinfo {author} {\bibfnamefont {P.~W.}\ \bibnamefont
  {Shor}},\ }\bibfield  {title} {\bibinfo {title} {Polynomial-time algorithms
  for prime factorization and discrete logarithms on a quantum computer},\
  }\href {https://doi.org/10.1137/S0097539795293172} {\bibfield  {journal}
  {\bibinfo  {journal} {SIAM review}\ }\textbf {\bibinfo {volume} {41}},\
  \bibinfo {pages} {303} (\bibinfo {year} {1999})}\BibitemShut {NoStop}%
\bibitem [{\citenamefont {Nielsen}\ and\ \citenamefont
  {Chuang}(2010)}]{NielsenChuang}%
  \BibitemOpen
  \bibfield  {author} {\bibinfo {author} {\bibfnamefont {M.~A.}\ \bibnamefont
  {Nielsen}}\ and\ \bibinfo {author} {\bibfnamefont {I.~L.}\ \bibnamefont
  {Chuang}},\ }\href {https://doi.org/10.1017/CBO9780511976667} {\emph
  {\bibinfo {title} {Quantum Computation and Quantum Information}}}\ (\bibinfo
  {publisher} {Cambridge University Press},\ \bibinfo {year}
  {2010})\BibitemShut {NoStop}%
\bibitem [{\citenamefont {Rivest}\ \emph {et~al.}(1978)\citenamefont {Rivest},
  \citenamefont {Shamir},\ and\ \citenamefont {Adleman}}]{rivest1978method}%
  \BibitemOpen
  \bibfield  {author} {\bibinfo {author} {\bibfnamefont {R.~L.}\ \bibnamefont
  {Rivest}}, \bibinfo {author} {\bibfnamefont {A.}~\bibnamefont {Shamir}},\
  and\ \bibinfo {author} {\bibfnamefont {L.}~\bibnamefont {Adleman}},\
  }\bibfield  {title} {\bibinfo {title} {A method for obtaining digital
  signatures and public-key cryptosystems},\ }\href
  {https://doi.org/10.1145/359340.359342} {\bibfield  {journal} {\bibinfo
  {journal} {Communications of the ACM}\ }\textbf {\bibinfo {volume} {21}},\
  \bibinfo {pages} {120} (\bibinfo {year} {1978})}\BibitemShut {NoStop}%
\bibitem [{\citenamefont {Haake}(2001)}]{Haake}%
  \BibitemOpen
  \bibfield  {author} {\bibinfo {author} {\bibfnamefont {F.}~\bibnamefont
  {Haake}},\ }\href {https://doi.org/10.1007/978-3-662-04506-0} {\emph
  {\bibinfo {title} {Quantum signatures of chaos}}}\ (\bibinfo  {publisher}
  {Springer, Berlin, Heidelberg},\ \bibinfo {year} {2001})\BibitemShut
  {NoStop}%
\bibitem [{\citenamefont {Ott}(2002)}]{Ott}%
  \BibitemOpen
  \bibfield  {author} {\bibinfo {author} {\bibfnamefont {E.}~\bibnamefont
  {Ott}},\ }\href {https://doi.org/10.1017/CBO9780511803260} {\emph {\bibinfo
  {title} {Chaos in dynamical systems}}}\ (\bibinfo  {publisher} {Cambridge
  University Press},\ \bibinfo {year} {2002})\BibitemShut {NoStop}%
\bibitem [{\citenamefont {McDonald}\ and\ \citenamefont
  {Kaufman}(1979)}]{DKchaos}%
  \BibitemOpen
  \bibfield  {author} {\bibinfo {author} {\bibfnamefont {S.~W.}\ \bibnamefont
  {McDonald}}\ and\ \bibinfo {author} {\bibfnamefont {A.~N.}\ \bibnamefont
  {Kaufman}},\ }\bibfield  {title} {\bibinfo {title} {Spectrum and
  eigenfunctions for a {Hamiltonian} with stochastic trajectories},\ }\href
  {https://doi.org/10.1103/PhysRevLett.42.1189} {\bibfield  {journal} {\bibinfo
   {journal} {Phys. Rev. Lett.}\ }\textbf {\bibinfo {volume} {42}},\ \bibinfo
  {pages} {1189} (\bibinfo {year} {1979})}\BibitemShut {NoStop}%
\bibitem [{\citenamefont {Casati}\ \emph {et~al.}(1980)\citenamefont {Casati},
  \citenamefont {Valz-Gris},\ and\ \citenamefont {Guarnieri}}]{CGV}%
  \BibitemOpen
  \bibfield  {author} {\bibinfo {author} {\bibfnamefont {G.}~\bibnamefont
  {Casati}}, \bibinfo {author} {\bibfnamefont {F.}~\bibnamefont {Valz-Gris}},\
  and\ \bibinfo {author} {\bibfnamefont {I.}~\bibnamefont {Guarnieri}},\
  }\bibfield  {title} {\bibinfo {title} {On the connection between quantization
  of nonintegrable systems and statistical theory of spectra},\ }\href
  {https://doi.org/10.1007/BF02798790} {\bibfield  {journal} {\bibinfo
  {journal} {Lett. Nuovo Cimento}\ }\textbf {\bibinfo {volume} {28}},\ \bibinfo
  {pages} {279} (\bibinfo {year} {1980})}\BibitemShut {NoStop}%
\bibitem [{\citenamefont {Berry}(1981)}]{BerryStadium}%
  \BibitemOpen
  \bibfield  {author} {\bibinfo {author} {\bibfnamefont {M.~V.}\ \bibnamefont
  {Berry}},\ }\bibfield  {title} {\bibinfo {title} {Quantizing a classically
  ergodic system: {Sinai}'s billiard and the {KKR} method},\ }\href
  {https://doi.org/10.1016/0003-4916(81)90189-5} {\bibfield  {journal}
  {\bibinfo  {journal} {Ann. Phys.}\ }\textbf {\bibinfo {volume} {131}},\
  \bibinfo {pages} {163} (\bibinfo {year} {1981})}\BibitemShut {NoStop}%
\bibitem [{\citenamefont {Bohigas}\ \emph {et~al.}(1984)\citenamefont
  {Bohigas}, \citenamefont {Giannoni},\ and\ \citenamefont {Schmit}}]{BGS}%
  \BibitemOpen
  \bibfield  {author} {\bibinfo {author} {\bibfnamefont {O.}~\bibnamefont
  {Bohigas}}, \bibinfo {author} {\bibfnamefont {M.-J.}\ \bibnamefont
  {Giannoni}},\ and\ \bibinfo {author} {\bibfnamefont {C.}~\bibnamefont
  {Schmit}},\ }\bibfield  {title} {\bibinfo {title} {Characterization of
  chaotic quantum spectra and universality of level fluctuation laws},\ }\href
  {https://doi.org/10.1103/PhysRevLett.52.1} {\bibfield  {journal} {\bibinfo
  {journal} {Phys. Rev. Lett.}\ }\textbf {\bibinfo {volume} {52}},\ \bibinfo
  {pages} {1} (\bibinfo {year} {1984})}\BibitemShut {NoStop}%
\bibitem [{\citenamefont {Hannay}\ and\ \citenamefont {Ozorio~de
  Almeida}(1984)}]{HOdA}%
  \BibitemOpen
  \bibfield  {author} {\bibinfo {author} {\bibfnamefont {J.~H.}\ \bibnamefont
  {Hannay}}\ and\ \bibinfo {author} {\bibfnamefont {A.~M.}\ \bibnamefont
  {Ozorio~de Almeida}},\ }\bibfield  {title} {\bibinfo {title} {Periodic orbits
  and a correlation function for the semiclassical density of states},\ }\href
  {https://doi.org/10.1088/0305-4470/17/18/013} {\bibfield  {journal} {\bibinfo
   {journal} {J. Phys. A: Math. Gen.}\ }\textbf {\bibinfo {volume} {17}},\
  \bibinfo {pages} {3429} (\bibinfo {year} {1984})}\BibitemShut {NoStop}%
\bibitem [{\citenamefont {Berry}(1985)}]{BerrySR}%
  \BibitemOpen
  \bibfield  {author} {\bibinfo {author} {\bibfnamefont {M.~V.}\ \bibnamefont
  {Berry}},\ }\bibfield  {title} {\bibinfo {title} {Semiclassical theory of
  spectral rigidity},\ }\href {https://doi.org/10.1098/rspa.1985.0078}
  {\bibfield  {journal} {\bibinfo  {journal} {Proc. Roy. Soc. Lond. A}\
  }\textbf {\bibinfo {volume} {400}},\ \bibinfo {pages} {229} (\bibinfo {year}
  {1985})}\BibitemShut {NoStop}%
\bibitem [{\citenamefont {Argaman}\ \emph {et~al.}(1993)\citenamefont
  {Argaman}, \citenamefont {Imry},\ and\ \citenamefont {Smilansky}}]{argaman}%
  \BibitemOpen
  \bibfield  {author} {\bibinfo {author} {\bibfnamefont {N.}~\bibnamefont
  {Argaman}}, \bibinfo {author} {\bibfnamefont {Y.}~\bibnamefont {Imry}},\ and\
  \bibinfo {author} {\bibfnamefont {U.}~\bibnamefont {Smilansky}},\ }\bibfield
  {title} {\bibinfo {title} {Semiclassical analysis of spectral correlations in
  mesoscopic systems},\ }\href {https://doi.org/10.1103/PhysRevB.47.4440}
  {\bibfield  {journal} {\bibinfo  {journal} {Phys. Rev. B}\ }\textbf {\bibinfo
  {volume} {47}},\ \bibinfo {pages} {4440} (\bibinfo {year}
  {1993})}\BibitemShut {NoStop}%
\bibitem [{\citenamefont {M{\"u}ller}\ \emph {et~al.}(2004)\citenamefont
  {M{\"u}ller}, \citenamefont {Heusler}, \citenamefont {Braun}, \citenamefont
  {Haake},\ and\ \citenamefont {Altland}}]{HaakePO}%
  \BibitemOpen
  \bibfield  {author} {\bibinfo {author} {\bibfnamefont {S.}~\bibnamefont
  {M{\"u}ller}}, \bibinfo {author} {\bibfnamefont {S.}~\bibnamefont {Heusler}},
  \bibinfo {author} {\bibfnamefont {P.}~\bibnamefont {Braun}}, \bibinfo
  {author} {\bibfnamefont {F.}~\bibnamefont {Haake}},\ and\ \bibinfo {author}
  {\bibfnamefont {A.}~\bibnamefont {Altland}},\ }\bibfield  {title} {\bibinfo
  {title} {Semiclassical foundation of universality in quantum chaos},\ }\href
  {https://doi.org/10.1103/PhysRevLett.93.014103} {\bibfield  {journal}
  {\bibinfo  {journal} {Phys. Rev. Lett.}\ }\textbf {\bibinfo {volume} {93}},\
  \bibinfo {pages} {014103} (\bibinfo {year} {2004})}\BibitemShut {NoStop}%
\bibitem [{\citenamefont {M{\"u}ller}\ \emph {et~al.}(2005)\citenamefont
  {M{\"u}ller}, \citenamefont {Heusler}, \citenamefont {Braun}, \citenamefont
  {Haake},\ and\ \citenamefont {Altland}}]{HaakePO2}%
  \BibitemOpen
  \bibfield  {author} {\bibinfo {author} {\bibfnamefont {S.}~\bibnamefont
  {M{\"u}ller}}, \bibinfo {author} {\bibfnamefont {S.}~\bibnamefont {Heusler}},
  \bibinfo {author} {\bibfnamefont {P.}~\bibnamefont {Braun}}, \bibinfo
  {author} {\bibfnamefont {F.}~\bibnamefont {Haake}},\ and\ \bibinfo {author}
  {\bibfnamefont {A.}~\bibnamefont {Altland}},\ }\bibfield  {title} {\bibinfo
  {title} {Periodic-orbit theory of universality in quantum chaos},\ }\href
  {https://doi.org/10.1103/PhysRevE.72.046207} {\bibfield  {journal} {\bibinfo
  {journal} {Phys. Rev. E}\ }\textbf {\bibinfo {volume} {72}},\ \bibinfo
  {pages} {046207} (\bibinfo {year} {2005})}\BibitemShut {NoStop}%
\bibitem [{\citenamefont {Vikram}\ and\ \citenamefont
  {Galitski}(2022)}]{dynamicalqergodicity}%
  \BibitemOpen
  \bibfield  {author} {\bibinfo {author} {\bibfnamefont {A.}~\bibnamefont
  {Vikram}}\ and\ \bibinfo {author} {\bibfnamefont {V.}~\bibnamefont
  {Galitski}},\ }\bibfield  {title} {\bibinfo {title} {Dynamical quantum
  ergodicity from energy level statistics},\ }\href
  {https://doi.org/10.48550/arXiv.2205.05704} {\bibfield  {journal} {\bibinfo
  {journal} {arXiv preprint arXiv:2205.05704}\ } (\bibinfo {year}
  {2022})}\BibitemShut {NoStop}%
\bibitem [{\citenamefont {Hannay}\ and\ \citenamefont
  {Berry}(1980)}]{hannay1980quantization}%
  \BibitemOpen
  \bibfield  {author} {\bibinfo {author} {\bibfnamefont {J.}~\bibnamefont
  {Hannay}}\ and\ \bibinfo {author} {\bibfnamefont {M.~V.}\ \bibnamefont
  {Berry}},\ }\bibfield  {title} {\bibinfo {title} {Quantization of linear maps
  on a torus-fresnel diffraction by a periodic grating},\ }\href
  {https://doi.org/10.1016/0167-2789(80)90026-3} {\bibfield  {journal}
  {\bibinfo  {journal} {Phys. D: Nonlinear Phenom.}\ }\textbf {\bibinfo
  {volume} {1}},\ \bibinfo {pages} {267} (\bibinfo {year} {1980})}\BibitemShut
  {NoStop}%
\bibitem [{\citenamefont {Keating}(1991)}]{keating1991cat}%
  \BibitemOpen
  \bibfield  {author} {\bibinfo {author} {\bibfnamefont {J.~P.}\ \bibnamefont
  {Keating}},\ }\bibfield  {title} {\bibinfo {title} {The cat maps: quantum
  mechanics and classical motion},\ }\href
  {https://doi.org/10.1088/0951-7715/4/2/006} {\bibfield  {journal} {\bibinfo
  {journal} {Nonlinearity}\ }\textbf {\bibinfo {volume} {4}},\ \bibinfo {pages}
  {309} (\bibinfo {year} {1991})}\BibitemShut {NoStop}%
\bibitem [{\citenamefont {Pako\'{n}ski}\ \emph {et~al.}(2001)\citenamefont
  {Pako\'{n}ski}, \citenamefont {\.{Z}yczkowski},\ and\ \citenamefont
  {Ku\'{s}}}]{QuantumGraphs1D}%
  \BibitemOpen
  \bibfield  {author} {\bibinfo {author} {\bibfnamefont {P.}~\bibnamefont
  {Pako\'{n}ski}}, \bibinfo {author} {\bibfnamefont {K.}~\bibnamefont
  {\.{Z}yczkowski}},\ and\ \bibinfo {author} {\bibfnamefont {M.}~\bibnamefont
  {Ku\'{s}}},\ }\bibfield  {title} {\bibinfo {title} {Classical {1D} maps,
  quantum graphs and ensembles of unitary matrices},\ }\href
  {https://doi.org/10.1088/0305-4470/34/43/313} {\bibfield  {journal} {\bibinfo
   {journal} {J. Phys. A: Math. Gen.}\ }\textbf {\bibinfo {volume} {34}},\
  \bibinfo {pages} {9303} (\bibinfo {year} {2001})}\BibitemShut {NoStop}%
\bibitem [{\citenamefont {Balazs}\ and\ \citenamefont
  {Voros}(1989)}]{balazs1989quantized}%
  \BibitemOpen
  \bibfield  {author} {\bibinfo {author} {\bibfnamefont {N.~L.}\ \bibnamefont
  {Balazs}}\ and\ \bibinfo {author} {\bibfnamefont {A.}~\bibnamefont {Voros}},\
  }\bibfield  {title} {\bibinfo {title} {The quantized baker's
  transformation},\ }\href {https://doi.org/10.1016/0003-4916(89)90259-5}
  {\bibfield  {journal} {\bibinfo  {journal} {Ann. Phys.}\ }\textbf {\bibinfo
  {volume} {190}},\ \bibinfo {pages} {1} (\bibinfo {year} {1989})}\BibitemShut
  {NoStop}%
\bibitem [{\citenamefont {Saraceno}(1990)}]{saraceno1990classical}%
  \BibitemOpen
  \bibfield  {author} {\bibinfo {author} {\bibfnamefont {M.}~\bibnamefont
  {Saraceno}},\ }\bibfield  {title} {\bibinfo {title} {Classical structures in
  the quantized baker transformation},\ }\href
  {https://doi.org/10.1016/0003-4916(90)90367-W} {\bibfield  {journal}
  {\bibinfo  {journal} {Ann. Phys.}\ }\textbf {\bibinfo {volume} {199}},\
  \bibinfo {pages} {37} (\bibinfo {year} {1990})}\BibitemShut {NoStop}%
\bibitem [{\citenamefont {Dyson}\ and\ \citenamefont {Falk}(1992)}]{dysoncat}%
  \BibitemOpen
  \bibfield  {author} {\bibinfo {author} {\bibfnamefont {F.~J.}\ \bibnamefont
  {Dyson}}\ and\ \bibinfo {author} {\bibfnamefont {H.}~\bibnamefont {Falk}},\
  }\bibfield  {title} {\bibinfo {title} {Period of a discrete cat mapping},\
  }\href {https://doi.org/10.2307/2324989} {\bibfield  {journal} {\bibinfo
  {journal} {The American Mathematical Monthly}\ }\textbf {\bibinfo {volume}
  {99}},\ \bibinfo {pages} {603} (\bibinfo {year} {1992})}\BibitemShut
  {NoStop}%
\bibitem [{\citenamefont {Lakshminarayan}(2005)}]{lakshminarayan2005shuffling}%
  \BibitemOpen
  \bibfield  {author} {\bibinfo {author} {\bibfnamefont {A.}~\bibnamefont
  {Lakshminarayan}},\ }\bibfield  {title} {\bibinfo {title} {Shuffling cards,
  factoring numbers and the quantum baker's map},\ }\href
  {https://doi.org/10.1088/0305-4470/38/37/L01} {\bibfield  {journal} {\bibinfo
   {journal} {J. Phys. A: Math. Gen.}\ }\textbf {\bibinfo {volume} {38}},\
  \bibinfo {pages} {L597} (\bibinfo {year} {2005})}\BibitemShut {NoStop}%
\bibitem [{\citenamefont {Lakshminarayan}(2007)}]{lakshminarayan2007modular}%
  \BibitemOpen
  \bibfield  {author} {\bibinfo {author} {\bibfnamefont {A.}~\bibnamefont
  {Lakshminarayan}},\ }\bibfield  {title} {\bibinfo {title} {Modular
  multiplication operator and quantized baker’s maps},\ }\href
  {https://doi.org/10.1103/PhysRevA.76.042330} {\bibfield  {journal} {\bibinfo
  {journal} {Phys. Rev. A}\ }\textbf {\bibinfo {volume} {76}},\ \bibinfo
  {pages} {042330} (\bibinfo {year} {2007})}\BibitemShut {NoStop}%
\bibitem [{\citenamefont {Balazs}\ and\ \citenamefont
  {Voros}(1987)}]{balazs1987quantized}%
  \BibitemOpen
  \bibfield  {author} {\bibinfo {author} {\bibfnamefont {N.}~\bibnamefont
  {Balazs}}\ and\ \bibinfo {author} {\bibfnamefont {A.}~\bibnamefont {Voros}},\
  }\bibfield  {title} {\bibinfo {title} {The quantized baker's
  transformation},\ }\href {https://doi.org/10.1209/0295-5075/4/10/001}
  {\bibfield  {journal} {\bibinfo  {journal} {Europhysics Letters}\ }\textbf
  {\bibinfo {volume} {4}},\ \bibinfo {pages} {1089} (\bibinfo {year}
  {1987})}\BibitemShut {NoStop}%
\bibitem [{\citenamefont {Anantharaman}\ and\ \citenamefont
  {Nonnenmacher}(2007)}]{AN}%
  \BibitemOpen
  \bibfield  {author} {\bibinfo {author} {\bibfnamefont {N.}~\bibnamefont
  {Anantharaman}}\ and\ \bibinfo {author} {\bibfnamefont {S.}~\bibnamefont
  {Nonnenmacher}},\ }\bibfield  {title} {\bibinfo {title} {Entropy of
  semiclassical measures of the {W}alsh-quantized baker's map},\ }\href
  {https://doi.org/10.1007/s00023-006-0299-z} {\bibfield  {journal} {\bibinfo
  {journal} {Ann. Henri Poincar\'{e}}\ }\textbf {\bibinfo {volume} {8}},\
  \bibinfo {pages} {37} (\bibinfo {year} {2007})}\BibitemShut {NoStop}%
\bibitem [{\citenamefont {R{\'e}nyi}(1957)}]{renyi1957representations}%
  \BibitemOpen
  \bibfield  {author} {\bibinfo {author} {\bibfnamefont {A.}~\bibnamefont
  {R{\'e}nyi}},\ }\bibfield  {title} {\bibinfo {title} {Representations for
  real numbers and their ergodic properties},\ }\href
  {https://doi.org/10.1007/BF02020331} {\bibfield  {journal} {\bibinfo
  {journal} {Acta Math. Acad. Sci. Hungar}\ }\textbf {\bibinfo {volume} {8}},\
  \bibinfo {pages} {477} (\bibinfo {year} {1957})}\BibitemShut {NoStop}%
\bibitem [{\citenamefont {Saraceno}\ and\ \citenamefont
  {Voros}(1994)}]{saraceno1994towards}%
  \BibitemOpen
  \bibfield  {author} {\bibinfo {author} {\bibfnamefont {M.}~\bibnamefont
  {Saraceno}}\ and\ \bibinfo {author} {\bibfnamefont {A.}~\bibnamefont
  {Voros}},\ }\bibfield  {title} {\bibinfo {title} {Towards a semiclassical
  theory of the quantum baker's map},\ }\href
  {https://doi.org/10.1016/S0167-2789(05)80007-7} {\bibfield  {journal}
  {\bibinfo  {journal} {Phys. D: Nonlinear Phenom.}\ }\textbf {\bibinfo
  {volume} {79}},\ \bibinfo {pages} {206} (\bibinfo {year} {1994})}\BibitemShut
  {NoStop}%
\bibitem [{\citenamefont {Rubin}\ and\ \citenamefont
  {Salwen}(1998)}]{rubin1998canonical}%
  \BibitemOpen
  \bibfield  {author} {\bibinfo {author} {\bibfnamefont {R.}~\bibnamefont
  {Rubin}}\ and\ \bibinfo {author} {\bibfnamefont {N.}~\bibnamefont {Salwen}},\
  }\bibfield  {title} {\bibinfo {title} {A canonical quantization of the
  baker's map},\ }\href {https://doi.org/10.1006/aphy.1998.5845} {\bibfield
  {journal} {\bibinfo  {journal} {Ann. Phys.}\ }\textbf {\bibinfo {volume}
  {269}},\ \bibinfo {pages} {159} (\bibinfo {year} {1998})}\BibitemShut
  {NoStop}%
\bibitem [{\citenamefont {Ermann}\ and\ \citenamefont
  {Saraceno}(2006)}]{ermann2006generalized}%
  \BibitemOpen
  \bibfield  {author} {\bibinfo {author} {\bibfnamefont {L.}~\bibnamefont
  {Ermann}}\ and\ \bibinfo {author} {\bibfnamefont {M.}~\bibnamefont
  {Saraceno}},\ }\bibfield  {title} {\bibinfo {title} {Generalized quantum
  baker maps as perturbations of a simple kernel},\ }\href
  {https://doi.org/10.1103/PhysRevE.74.046205} {\bibfield  {journal} {\bibinfo
  {journal} {Phys. Rev. E}\ }\textbf {\bibinfo {volume} {74}},\ \bibinfo
  {pages} {046205} (\bibinfo {year} {2006})}\BibitemShut {NoStop}%
\bibitem [{\citenamefont {De~Bi\`evre}\ and\ \citenamefont
  {Degli~Esposti}(1998)}]{DBDE}%
  \BibitemOpen
  \bibfield  {author} {\bibinfo {author} {\bibfnamefont {S.}~\bibnamefont
  {De~Bi\`evre}}\ and\ \bibinfo {author} {\bibfnamefont {M.}~\bibnamefont
  {Degli~Esposti}},\ }\bibfield  {title} {\bibinfo {title} {Egorov theorems and
  equidistribution of eigenfunctions for the quantized sawtooth and baker
  maps},\ }\href {http://www.numdam.org/item/AIHPA_1998__69_1_1_0/} {\bibfield
  {journal} {\bibinfo  {journal} {Ann. Inst. H. Poincar\'{e} Phys. Th\'{e}or.}\
  }\textbf {\bibinfo {volume} {69}},\ \bibinfo {pages} {1} (\bibinfo {year}
  {1998})}\BibitemShut {NoStop}%
\bibitem [{\citenamefont {Degli~Esposti}\ \emph {et~al.}(2006)\citenamefont
  {Degli~Esposti}, \citenamefont {Nonnenmacher},\ and\ \citenamefont
  {Winn}}]{DNW}%
  \BibitemOpen
  \bibfield  {author} {\bibinfo {author} {\bibfnamefont {M.}~\bibnamefont
  {Degli~Esposti}}, \bibinfo {author} {\bibfnamefont {S.}~\bibnamefont
  {Nonnenmacher}},\ and\ \bibinfo {author} {\bibfnamefont {B.}~\bibnamefont
  {Winn}},\ }\bibfield  {title} {\bibinfo {title} {Quantum variance and
  ergodicity for the baker's map},\ }\href
  {https://doi.org/10.1007/s00220-005-1397-3} {\bibfield  {journal} {\bibinfo
  {journal} {Comm. Math. Phys.}\ }\textbf {\bibinfo {volume} {263}},\ \bibinfo
  {pages} {325} (\bibinfo {year} {2006})}\BibitemShut {NoStop}%
\bibitem [{\citenamefont {Tracy}\ and\ \citenamefont
  {Scott}(2002)}]{TracyScott}%
  \BibitemOpen
  \bibfield  {author} {\bibinfo {author} {\bibfnamefont {M.~M.}\ \bibnamefont
  {Tracy}}\ and\ \bibinfo {author} {\bibfnamefont {A.~J.}\ \bibnamefont
  {Scott}},\ }\bibfield  {title} {\bibinfo {title} {The classical limit for a
  class of quantum baker's maps},\ }\href
  {https://doi.org/10.1088/0305-4470/35/39/314} {\bibfield  {journal} {\bibinfo
   {journal} {J. Phys. A}\ }\textbf {\bibinfo {volume} {35}},\ \bibinfo {pages}
  {8341} (\bibinfo {year} {2002})}\BibitemShut {NoStop}%
\bibitem [{\citenamefont {Brun}\ and\ \citenamefont
  {Schack}(1999)}]{bakingprotocol}%
  \BibitemOpen
  \bibfield  {author} {\bibinfo {author} {\bibfnamefont {T.~A.}\ \bibnamefont
  {Brun}}\ and\ \bibinfo {author} {\bibfnamefont {R.}~\bibnamefont {Schack}},\
  }\bibfield  {title} {\bibinfo {title} {Realizing the quantum baker’s map on
  a {NMR} quantum computer},\ }\href {https://doi.org/10.1103/PhysRevA.59.2649}
  {\bibfield  {journal} {\bibinfo  {journal} {Phys. Rev. A}\ }\textbf {\bibinfo
  {volume} {59}},\ \bibinfo {pages} {2649} (\bibinfo {year}
  {1999})}\BibitemShut {NoStop}%
\bibitem [{\citenamefont {Weinstein}\ \emph {et~al.}(2002)\citenamefont
  {Weinstein}, \citenamefont {Lloyd}, \citenamefont {Emerson},\ and\
  \citenamefont {Cory}}]{bakingsimulationNMR}%
  \BibitemOpen
  \bibfield  {author} {\bibinfo {author} {\bibfnamefont {Y.~S.}\ \bibnamefont
  {Weinstein}}, \bibinfo {author} {\bibfnamefont {S.}~\bibnamefont {Lloyd}},
  \bibinfo {author} {\bibfnamefont {J.}~\bibnamefont {Emerson}},\ and\ \bibinfo
  {author} {\bibfnamefont {D.~G.}\ \bibnamefont {Cory}},\ }\bibfield  {title}
  {\bibinfo {title} {Experimental implementation of the quantum baker’s
  map},\ }\href {https://doi.org/https://doi.org/10.1103/PhysRevLett.89.157902}
  {\bibfield  {journal} {\bibinfo  {journal} {Phys. Rev. Lett.}\ }\textbf
  {\bibinfo {volume} {89}},\ \bibinfo {pages} {157902} (\bibinfo {year}
  {2002})}\BibitemShut {NoStop}%
\bibitem [{\citenamefont {Vasilyev}\ \emph {et~al.}(2020)\citenamefont
  {Vasilyev}, \citenamefont {Grankin}, \citenamefont {Baranov}, \citenamefont
  {Sieberer},\ and\ \citenamefont {Zoller}}]{QNDSFF}%
  \BibitemOpen
  \bibfield  {author} {\bibinfo {author} {\bibfnamefont {D.~V.}\ \bibnamefont
  {Vasilyev}}, \bibinfo {author} {\bibfnamefont {A.}~\bibnamefont {Grankin}},
  \bibinfo {author} {\bibfnamefont {M.~A.}\ \bibnamefont {Baranov}}, \bibinfo
  {author} {\bibfnamefont {L.~M.}\ \bibnamefont {Sieberer}},\ and\ \bibinfo
  {author} {\bibfnamefont {P.}~\bibnamefont {Zoller}},\ }\bibfield  {title}
  {\bibinfo {title} {Monitoring quantum simulators via quantum nondemolition
  couplings to atomic clock qubits},\ }\href
  {https://doi.org/10.1103/PRXQuantum.1.020302} {\bibfield  {journal} {\bibinfo
   {journal} {PRX Quantum}\ }\textbf {\bibinfo {volume} {1}},\ \bibinfo {pages}
  {020302} (\bibinfo {year} {2020})}\BibitemShut {NoStop}%
\bibitem [{\citenamefont {Joshi}\ \emph {et~al.}(2022)\citenamefont {Joshi},
  \citenamefont {Elben}, \citenamefont {Vikram}, \citenamefont {Vermersch},
  \citenamefont {Galitski},\ and\ \citenamefont {Zoller}}]{pSFF}%
  \BibitemOpen
  \bibfield  {author} {\bibinfo {author} {\bibfnamefont {L.~K.}\ \bibnamefont
  {Joshi}}, \bibinfo {author} {\bibfnamefont {A.}~\bibnamefont {Elben}},
  \bibinfo {author} {\bibfnamefont {A.}~\bibnamefont {Vikram}}, \bibinfo
  {author} {\bibfnamefont {B.}~\bibnamefont {Vermersch}}, \bibinfo {author}
  {\bibfnamefont {V.}~\bibnamefont {Galitski}},\ and\ \bibinfo {author}
  {\bibfnamefont {P.}~\bibnamefont {Zoller}},\ }\bibfield  {title} {\bibinfo
  {title} {Probing many-body quantum chaos with quantum simulators},\ }\href
  {https://doi.org/10.1103/PhysRevX.12.011018} {\bibfield  {journal} {\bibinfo
  {journal} {Phys. Rev. X}\ }\textbf {\bibinfo {volume} {12}},\ \bibinfo
  {pages} {011018} (\bibinfo {year} {2022})}\BibitemShut {NoStop}%
\bibitem [{\citenamefont {Bouzouina}\ and\ \citenamefont
  {De~Bi\`evre}(1996)}]{bdb}%
  \BibitemOpen
  \bibfield  {author} {\bibinfo {author} {\bibfnamefont {A.}~\bibnamefont
  {Bouzouina}}\ and\ \bibinfo {author} {\bibfnamefont {S.}~\bibnamefont
  {De~Bi\`evre}},\ }\bibfield  {title} {\bibinfo {title} {Equipartition of the
  eigenfunctions of quantized ergodic maps on the torus},\ }\href
  {https://doi.org/10.1007/BF02104909} {\bibfield  {journal} {\bibinfo
  {journal} {Comm. Math. Phys.}\ }\textbf {\bibinfo {volume} {178}},\ \bibinfo
  {pages} {83} (\bibinfo {year} {1996})}\BibitemShut {NoStop}%
\end{thebibliography}%

\pagebreak
\widetext
\begin{center}
\textbf{\large Supplementary Material for\\ ``Chaotic Roots of the Modular Multiplication Dynamical System in Shor's Algorithm''}
\end{center}
\setcounter{equation}{0}
\setcounter{figure}{0}
\setcounter{table}{0}
\setcounter{page}{1}
\makeatletter
\renewcommand{\theequation}{S\arabic{equation}}
\renewcommand{\thefigure}{S\arabic{figure}}

\section{Quantization of the A-baker's map}\label{sec:detail}
In this section, we will explain the quantization procedure of the A-baker's map based on the work by Balazs and Voros~\cite{balazs1989quantized}. Classically A-baker's map transforms a unit square phase space onto itself according to the following equations: 
\begin{align} 
    x \rightarrow x^{'} = Ax - \lfloor Ax \rfloor, \\ 
    p \rightarrow p^{'} = \frac{p+\lfloor Ax \rfloor}{A}.
\end{align}
Here $x$ and $p$ are two variables of phase space and $A$ is an integer. 
\begin{figure}[htb] 
    \centering
    \includegraphics[width=.80\columnwidth]{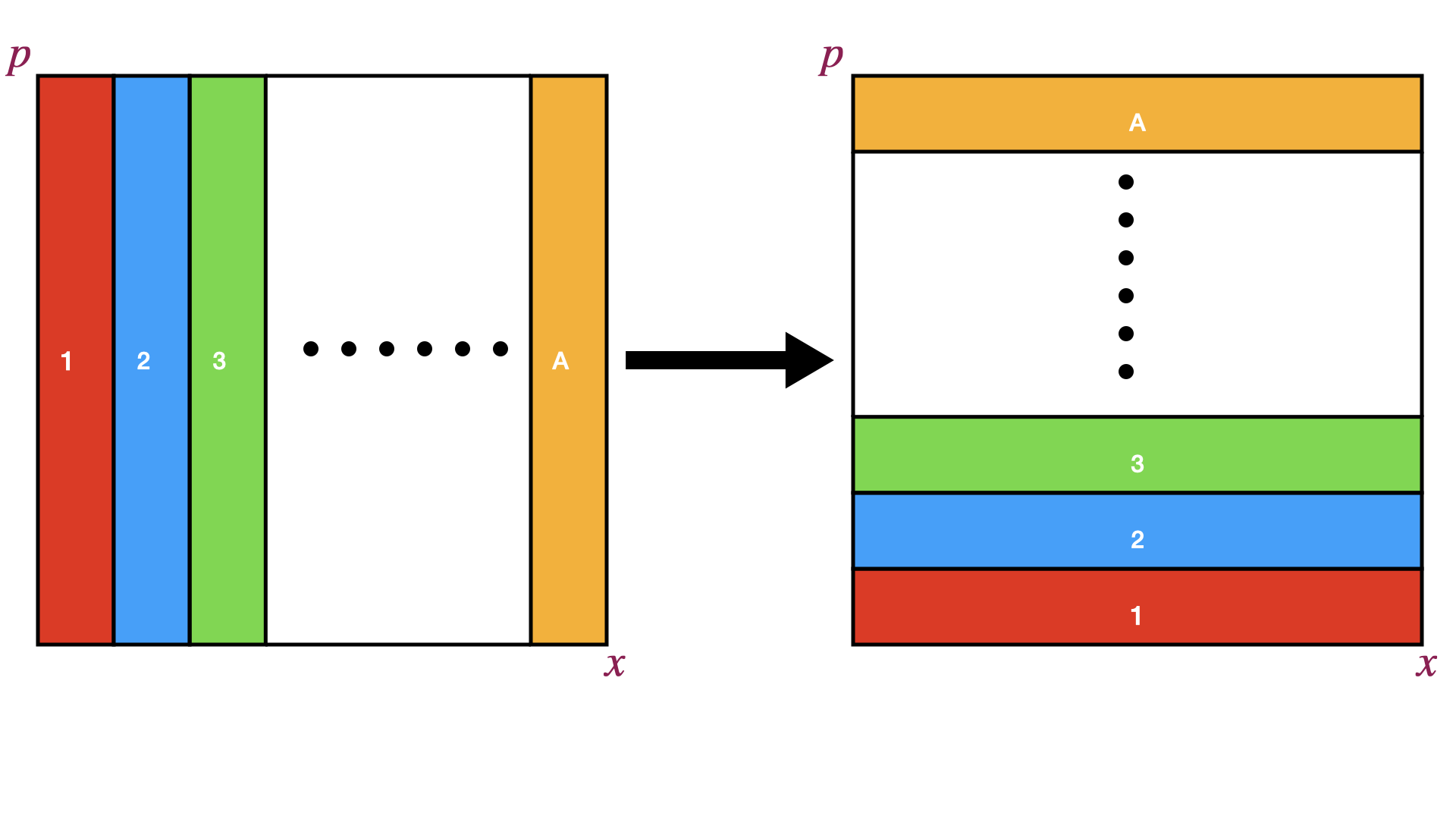}
    \caption{A-baker's map transforms the unit square in the left to the one in the right. The unit square is divided into $A$ rectangles which are marked by numbers from $1$ to $A$. Each rectangle is stretched along $x$ axis and contracted along $p$ axis before being stacked on top of each other.}
    \label{fig:s1}
\end{figure}
There is a simple pictorial representation of the map which is shown in Fig. \ref{fig:s1}.
First divide the unit square into $A$ rectangles. Then stretch each rectangles in horizontal direction and contracts in vertical direction and finally stack the rectangles on top of each other. To quantize this map, instead of a phase space we consider a $D$ dimensional Hilbert space. Suppose $\{\ket{x_n}\}_{n=0}^{D-1}$ and $\{\ket{p_m}\}_{m=0}^{D-1}$ are two different complete basis. We assume that they satisfy the following boundary conditions:
\begin{align} \label{supp_BC of q}
    \ket{x_{n+D}}& =e^{-2\pi i \beta} \ket{x_n},\\ \label{supp_BC of p}
    \ket{p_{n+D}}& =e^{2\pi i \alpha} \ket{p_n}.
\end{align}
When $\alpha=\beta=0$ it becomes periodic boundary condition. Physically one can think of these two basis as the eigenbasis of discrete position and momentum operator. The inner product of them is
\begin{equation}\label{supp_inner product of q&p}
    \bra{p_m}\ket{x_n} = \frac{1}{\sqrt{D}}e^{-\frac{2\pi i (n+\alpha) (m+\beta)}{D}}=[F_D^{\alpha, \beta}]_{nm},
\end{equation}
where $F_D^{\alpha, \beta}$ is a $D\cross D$ DFT matrix with phase $\alpha$ and $\beta$. Let's look at an arbitrary state $\ket{\psi}$ in the Hilbert space. We write the state as
\begin{align} \label{supp_dividing states in position}
    \ket{\psi} &= \sum_{n=0}^{D-1}\ket{x_n}\bra{x_n}\ket{\psi} \nonumber \\
               &= \sum_{n=0}^{\frac{D}{A}-1} \ket{x_n}\bra{x_n}\ket{\psi} + \sum_{n=\frac{D}{A}}^{\frac{2D}{A}-1} \ket{x_n}\bra{x_n}\ket{\psi} \dots \sum_{n=\frac{D(A-1)}{A}}^{D-1} \ket{x_n}\bra{x_n}\ket{\psi} \nonumber \\
               &= \sum_{l=1}^{A}\ket{\psi_l}, 
\end{align}
where $\ket{\psi_l}=\sum_{n=\frac{D(l-1)}{A}}^{\frac{Dl}{A}-1}\ket{x_n}\bra{x_n}\ket{\psi}$. Essentially we have partitioned the state in such a way that the projection of each partition into position basis can have non-zero value only inside a region. The regions are obtained from the equation
\begin{align} \label{supp_condition of psil}  
\bra{x_n}\ket{\psi_l}&=0 \hspace{.25 cm}\text{for} \hspace{.25 cm}0 \leq n \leq \frac{D(l-1)}{A}-1 \hspace{.25 cm} \text{and} \hspace{.25 cm} \frac{Dl}{A} \leq n \leq D-1.
\end{align}
The quantized A-baker's map is the unitary operator $B_A^{(0)}$ such that 
\begin{align} \label{supp_quantized Bernoulli} 
B_A^{(0)} \ket{\psi_l} &= \ket{\phi_l} \hspace{.25 cm} \text{for} \hspace{.25 cm} l=1,2 \dots A.
\end{align}
Eq. \eqref{supp_quantized Bernoulli} implies that $B_A^{(0)} \ket{\psi} = \sum_{l=1}^{A}\ket{\phi_l} \equiv \ket{\phi}$. The states $\ket{\phi_l}$ are defined in a similar way as Eq. \eqref{supp_dividing states in position} but in momentum basis i.e. $\ket{\phi_l}=\sum_{n=\frac{D(l-1)}{A}}^{\frac{Dl}{A}-1}\ket{p_n}\bra{p_n}\ket{\phi}$. Furthermore $\ket{\phi_l}$ satisfies
\begin{align}
    \bra{x_{An}}\ket{\phi_l}&= \frac{1}{\sqrt{A}} \bra{x_n}\ket{\psi_l} \hspace{.3 cm} \text{for} \hspace{.3 cm} \frac{D(l-1)}{A}\leq n \leq \frac{Dl}{A}-1, \label{supp_expansion along q}\\
    \bra{p_{n}}\ket{\phi_l}&= \sqrt{A} \bra{p_{An}}\ket{\psi_l} \hspace{.2 cm} \text{for} \hspace{.3 cm} \frac{D(l-1)}{A}\leq n \leq \frac{Dl}{A}-1 \label{supp_contraction along p}.
\end{align}
Eqs. \eqref{supp_expansion along q} and \eqref{supp_contraction along p} are analogous to expansion and contraction in the classical map. We insert a resolution of identity in Eq. \eqref{supp_contraction along p} and use the inner product of $\ket{x_n}$ and $\ket{p_m}$  to write
\begin{align}
    \bra{p_{n}}\ket{\phi_l} &= \sqrt{A} \sum_{m=0}^{D-1} \bra{p_{An}}\ket{x_m}\bra{x_m}\ket{\psi_l} \hspace{1.5 cm} \text{for} \hspace{.3 cm} \frac{D(l-1)}{A}\leq n \leq \frac{Dl}{A}-1 \nonumber \\
                            &= \sqrt{\frac{A}{D}} \sum_{m=0}^{D-1} e^{-\frac{2\pi i A(n+\alpha) (m+\beta)}{D}} \bra{x_m}\ket{\psi_l} \nonumber \\ \label{supp_expression of B_A0}
                            &=  \sum_{m=\frac{D(l-1)}{A}}^{\frac{Dl}{A}-1} (F_{\frac{D}{A}}^{\alpha, \beta})_{nm}\bra{x_m}\ket{\psi_l}. \hspace{1.5 cm} \text{[using Eq. \eqref{supp_condition of psil}]}
\end{align}
Note that $\ket{\phi_l}$ is defined in such a way that $\bra{p_{n}}\ket{\phi_l}$ when $\leq n \leq \frac{D(l-1)}{A}-1$ and $\frac{Dl}{A} \leq n \leq D-1.$ Combining it with Eq. \eqref{supp_expression of B_A0} we get a relation between $\ket{\phi_l}$ and $\ket{\psi_l}$. However, in Eq. \eqref{supp_expression of B_A0} they are in different representation. One can show that $\bra{p_{n}}\ket{\phi_l} = \sum_{m=0}^{D-1} (F_D^{\alpha, \beta})_{mn} \bra{x_{m}}\ket{\phi_l}$. Consequently we can rewrite Eq. \eqref{supp_expression of B_A0} as
\begin{equation} \label{supp_B_A0 in position basis}
    \bra{x_{n}}\ket{\phi_l}= \sum_{k=0}^{D-1} \sum_{m=\frac{D(l-1)}{A}}^{\frac{Dl}{A}-1}(F_D^{\alpha, \beta})_{nk}^{-1} (F_{\frac{D}{A}}^{\alpha, \beta})_{km}\bra{x_m}\ket{\psi_l}.
\end{equation}
We compare Eq. \eqref{supp_B_A0 in position basis} with Eq. \eqref{supp_quantized Bernoulli} to get an expression for quantized A-baker's map $B_A^{(0)}$:
\begin{equation} \label{supp_B_A0 matrix form}
    B_A^{(0)} = (F_D^{\alpha, \beta})^{-1} \begin{bmatrix}
        F_{\frac{D}{A}}^{\alpha, \beta} & 0 & 0 & \dots & 0 \\
        0 & F_{\frac{D}{A}}^{\alpha, \beta} & 0 & \dots & 0 \\
        0 & 0 & F_{\frac{D}{A}}^{\alpha, \beta} & \dots & 0 \\
        \vdots & \vdots & \vdots & \ddots & \vdots \\
        0 & 0 & 0 & \dots & F_{\frac{D}{A}}^{\alpha, \beta}
    \end{bmatrix}.
\end{equation}
We have derived the matrix form of $B_A^0$ without using Eq. \eqref{supp_expansion along q} . But one can check that $B_A^{(0)}$ is consistent with Eq. \eqref{supp_expansion along q}.  
\section{Analytical derivation of the relation between modular multiplication and the A-baker's map}\label{sec:analytic}
In this section we will derive Eq. \eqref{mod mult in terms of Bernoulli} of the main paper. According to this equation, if $N$ and $A$ are two co-primes such that $N=qA \pm 1$ then modular multiplication matrix $U_A^{\pm}$ satisfies
\begin{equation} \label{supp_mod mult in terms of DFT}
    U_A^{\pm} = F_{Aq}^{-1} \begin{bmatrix} F_q & F_q^{0,\mp \frac{1}{A}} & F_q^{0,\mp \frac{2}{A}} & \dots & F_q^{0, \mp \frac{A-1}{A}}\\ F_q &  F_q^{0,\mp \frac{1}{A}} &  F_q^{0,\mp \frac{2}{A}} & \dots &  F_q^{0,\mp \frac{A-1}{A}} \\ F_q &  F_q^{0,\mp \frac{1}{A}} &  F_q^{0,\mp \frac{2}{A}} & \dots &  F_q^{0,\mp \frac{A-1}{A}} \\ \vdots & \vdots & \vdots & \vdots & \vdots \\ F_q &  F_q^{0,\mp \frac{1}{A}} &  F_q^{0,\mp \frac{2}{A}} & \dots &  F_q^{0,\mp \frac{A-1}{A}}\end{bmatrix}   \odot [F_A^{\mp} \otimes J_q].
\end{equation}
Here $J_q$ is a $q \cross q$ matrix whose all elements are $1$, $F_A$ is the $A \cross A$ DFT matrix with no phase, $\otimes$ denotes tensor product and $\odot$ denotes Hadamard or entry-wise product of two matrices defined as $(A\odot B)_{ij} = A_{ij}B_{ij}$.
The derivation is slightly different for two cases of $N=qA \pm 1$. We will do the calculation for $N=Aq+1$ and point out the differences with $N=Aq-1$ case. The modular multiplication is defined as
\begin{equation} \label{supp_definition: Mod Multiplication}
U_A\ket{m}=\ket{mA \pmod{N}} \text{for \hspace{.1 cm}} m = 0,1 \dots ,N-1.
\end{equation}
From the definition one can see that the state $\ket{0}$ is always a fixed point of the modular multiplication. When $N=Aq+1$ we remove the $\ket{0}$ state such that the number of remaining states is a multiple of $A$. We denote this operator of dimension $Aq \times Aq$ as $U^{+}$. Using Eq. \eqref{supp_definition: Mod Multiplication} we find that
\begin{equation} \label{supp_modmult as Kronecker delta}
    \bra{n}U_A^{+}\ket{m} = \delta_{n,Am-lN} \hspace{.3 cm} \text{for} \hspace{.3 cm} lq+1\leq m \leq (l+1)q \hspace{.3 cm} \text{and} \hspace{.3 cm} l = 0,1 \dots ,A-1.
\end{equation}
Here $\delta_{n,m}$ denotes Kronecker delta function. Since we have removed the $\ket{0}$ state all the blocks with same $l$ have the same size. The Kronecker delta function can be written as
\begin{align}
    \delta_{n,Am-lN} &= \frac{1}{Aq}\sum_{k=1}^{Aq}e^{\frac{2\pi i k (n-Am+lN)}{Aq}} \nonumber \\
                  &= \frac{1}{Aq}\sum_{k=1}^{Aq}e^{\frac{2\pi i k n}{Aq}}e^{\frac{-2\pi i k (m-\frac{l}{A})}{q}} \nonumber \\
                  &=\frac{1}{Aq}[\sum_{k=1}^{q}e^{\frac{2\pi i k n}{Aq}}e^{\frac{-2\pi i k (m-\frac{l}{A})}{q}}+\sum_{k=q+1}^{2q}e^{\frac{2\pi i k n}{Aq}}e^{\frac{-2\pi i k (m-\frac{l}{A})}{q}} 
                  \dots +\sum_{k=(A-1)q+1}^{Aq}e^{\frac{2\pi i k n}{Aq}}e^{\frac{-2\pi i k (m-\frac{l}{A})}{q}}] \nonumber \\
                  &=\frac{1}{Aq}[\sum_{k=1}^{q}e^{\frac{2\pi i k n}{Aq}}e^{\frac{-2\pi i k (m-\frac{l}{A})}{q}}+\sum_{k=1}^{q}e^{\frac{2\pi i (k+q) n}{Aq}}e^{\frac{-2\pi i k (m-\frac{l}{A})}{q}}e^{\frac{-2\pi i l}{A}}
                   \dots +\sum_{k=1}^{q}e^{\frac{2\pi i [k+(A-1)q] n}{Aq}}e^{\frac{-2\pi i k (m-\frac{l}{A})}{q}}e^{\frac{-2\pi i l(A-1)}{A}}] \nonumber \\  \label{supp_delta in terms of DFT}
                  & = \frac{1}{\sqrt{A}}\sum_{k=1}^q[(F_{Aq})_{nk}^{-1}(F_q^{0,-\frac{l}{A}})_{km}+(F_{Aq})_{n(k+q)}^{-1}\omega_A^l(F_q^{0,-\frac{l}{A}})_{km}  
                   \dots +(F_{Aq})_{n[k+(A-1)q]}^{-1}\omega_A^{l(A-1)}(F_q^{0,-\frac{l}{A}})_{km}].
\end{align}
Here $\omega_A = e^{\frac{2\pi i}{A}}$ is the $A^{th}$ root of unity. We have expressed  the elements of modular multiplication in terms of DFT matrices. Using Eq. \eqref{supp_delta in terms of DFT} we can write modular multiplication as
\begin{equation} \label{supp_U in terms of DFT}
U_A^{+} = \frac{1}{\sqrt{A}}F_{Aq}^{-1} \begin{bmatrix} F_q & F_q^{0,- \frac{1}{A}} & F_q^{0,- \frac{2}{A}} & \dots & F_q^{0, - \frac{A-1}{A}}\\ F_q & \omega_A^{1} F_q^{0,- \frac{1}{A}} & \omega_A^{ 2} F_q^{0,- \frac{2}{A}} & \dots & \omega_A^{ A-1} F_q^{0,- \frac{A-1}{A}} \\ F_q & \omega_A^{ 2} F_q^{0,- \frac{1}{A}} & \omega_A^{ 4} F_q^{0,- \frac{2}{A}} & \dots & \omega_A^{ 2(A-1)} F_q^{0,- \frac{A-1}{A}} \\ \vdots & \vdots & \vdots & \vdots & \vdots \\ F_q & \omega_A^{ A-1} F_q^{0,- \frac{1}{A}} & \omega_A^{ 2(A-1)} F_q^{0,- \frac{2}{A}} & \dots & \omega_A^{ (A-1)^2} F_q^{0,- \frac{A-1}{A}}\end{bmatrix}.  \end{equation}
We can utilize Hadamard product to rewrite Eq. \eqref{supp_U in terms of DFT} as 
\begin{align}
    U_A^{+} & = F_{Aq}^{-1} \begin{bmatrix} F_q & F_q^{0,- \frac{1}{A}} & F_q^{0,- \frac{2}{A}} & \dots & F_q^{0, - \frac{A-1}{A}}\\ F_q &  F_q^{0,- \frac{1}{A}} &  F_q^{0,- \frac{2}{A}} & \dots &  F_q^{0,- \frac{A-1}{A}} \\ F_q &  F_q^{0,- \frac{1}{A}} &  F_q^{0,- \frac{2}{A}} & \dots &  F_q^{0,- \frac{A-1}{A}} \\ \vdots & \vdots & \vdots & \vdots & \vdots \\ F_q & F_q^{0,- \frac{1}{A}} &  F_q^{0,- \frac{2}{A}} & \dots &  F_q^{0,- \frac{A-1}{A}}\end{bmatrix} \odot \frac{1}{\sqrt{A}} \begin{bmatrix}
        \omega_A^{0} & \omega_A^{0} & \omega_A^{0} & \dots & \omega_A^{0}\\ \omega_A^{0} & \omega_A & \omega_A^{2} & \dots & \omega_A^{A-1}\\ \omega_A^{0} & \omega_A^{2} & \omega_A^{4} & \dots & \omega_A^{2(A-1)}\\ \vdots & \vdots & \vdots & \vdots & \vdots \\ \omega_A^{0} & \omega_A^{A-1} & \omega_A^{2(A-1)} & \dots & \omega_A^{(A-1)^2}
    \end{bmatrix}\otimes J_q \nonumber \\
        & = F_{Aq}^{-1} \begin{bmatrix} F_q & F_q^{0,- \frac{1}{A}} & F_q^{0,- \frac{2}{A}} & \dots & F_q^{0, - \frac{A-1}{A}}\\ F_q &  F_q^{0,- \frac{1}{A}} &  F_q^{0,- \frac{2}{A}} & \dots &  F_q^{0,- \frac{A-1}{A}} \\ F_q &  F_q^{0,- \frac{1}{A}} &  F_q^{0,- \frac{2}{A}} & \dots &  F_q^{0,- \frac{A-1}{A}} \\ \vdots & \vdots & \vdots & \vdots & \vdots \\ F_q & F_q^{0,- \frac{1}{A}} &  F_q^{0,- \frac{2}{A}} & \dots &  F_q^{0,- \frac{A-1}{A}}\end{bmatrix} \odot [F_A^{-1}\otimes J_q] .
\end{align}
When $N=Aq-1$ we add the state $\ket{N}$, which is a fixed state, so that the total number of states is a multiple of $A$. After adding the extra state we denote the operator as $U^{-}$ and perform a similar calculation as above to find
\begin{equation}
     U_A^{-} = F_{Aq}^{-1} \begin{bmatrix} F_q & F_q^{0,\frac{1}{A}} & F_q^{0, \frac{2}{A}} & \dots & F_q^{0,  \frac{A-1}{A}}\\ F_q &  F_q^{0, \frac{1}{A}} &  F_q^{0,\frac{2}{A}} & \dots &  F_q^{0,\frac{A-1}{A}} \\ F_q &  F_q^{0, \frac{1}{A}} &  F_q^{0,\frac{2}{A}} & \dots &  F_q^{0,\frac{A-1}{A}} \\ \vdots & \vdots & \vdots & \vdots & \vdots \\ F_q & F_q^{0,\frac{1}{A}} &  F_q^{0,\frac{2}{A}} & \dots &  F_q^{0, \frac{A-1}{A}}\end{bmatrix} \odot [F_A \otimes J_q].
\end{equation}

\section{Semiclassical limit}\label{sec:limit}

In this section we consider the action of the matrices $\tilde{B}_A^{\pm(k)}$ defined in Eq.~\eqref{Bernoulli tilde} on states in the semiclassical limit $D\to\infty$. For $\z_0=(x_0,p_0)$ in the torus $\T^2=\R^2/\Z^2$ (which we identify with the square $[0,1)\times[0,1)$) away from the discontinuities of the classical $A$-baker map, we will show that $\tilde{B}_A^{\pm(k)}$ evolves a Gaussian state localized at $\z_0$ to one localized close to the classical evolution value $S_A^{(k)}(\z_0):=(Ax-\lfloor Ax\rfloor,\frac{p+(\lfloor Ax\rfloor-k)\;\mathrm{mod}A}{A})$.
For this reason we refer to the matrices $\tilde{B}_A^{\pm(k)}$ as a quantum version of the classical $A$-baker map (with a cyclic $k$-shift).
We note that due to the phases $\alpha_j,\beta_j$, the quantum evolved state is actually centered at a point $S_A^{(k)}(\z_0)+\mathcal{O}_A(D^{-1})$, 
but this will end up being sufficiently close to the state centered at $S_A^{(k)}(\z_0)$ in the limit $D\to\infty$ for our purposes here. 
This error slows down the rate of convergence as $D\to\infty$ however.

In this section, $D$ is the dimension of the matrices, and we will allow $A$ to grow with the dimension $D$ as long as $A=o(D^{1/2}(\log D)^{-1/2})$. 
The computations here will be very similar to those in \cite[Prop. 5]{DNW}, though the addition of the phases $\alpha$ and $\beta$ in the generalized DFT matrices will require some additional consideration, e.g. Eq.~\eqref{eqn:cstate-bound}.

Besides cyclic shifts by $k$ elements, we will allow any permutation of blocks in $\tilde{B}_A^{+(0)}$ in Eq.~\eqref{Bernoulli tilde}. 
To this end, let $\Pi:\{0,\ldots,A-1\}\to\{0,\ldots,A-1\}$ be a permutation on $A$ elements, and let $P_\Pi$ be the $A\times A$ permutation matrix corresponding to $\Pi$. To make this permutation act on blocks, define the $D\times D$ matrix $P_{D,\Pi}:=P_\Pi\otimes I_{D/A}$, where $I_{D/A}$ is the $D/A\times D/A$ identity matrix.
Then for $D\in A\N$, define
\begin{align}
\tilde{B}_A^{(\Pi)}&:=F_D^{-1}\,P_{D,\Pi}\begin{bmatrix} \omega_0 F_{D/A}^{\alpha_0,\beta_0} &0&\cdots& 0\\
0&\omega_1 F_{D/A}^{\alpha_1,\beta_1} &\cdots&0\\
\vdots&&\ddots &\vdots\\
0&\cdots&0&\omega_{A-1} F_{D/A}^{\alpha_{A-1},\beta_{A-1}}\end{bmatrix},
\end{align}
where $\omega_j$ are arbitrary phase factors, $\alpha_j,\beta_j\in[-1,1]$, and $P_{D,\Pi}$ acts by permuting row blocks on the block matrix to the right according to $\Pi$.
In this notation, $\tilde{B}_A^{\pm(k)}$ from the main text corresponds to the special case of $\tilde{B}_A^{(\Pi)}$, where $\Pi$ is the cyclic shift by $k$ elements $\Pi^{(k)}:n\mapsto n-k\;\mathrm{mod}\,A$, $\alpha_j=0$,  $\beta_j=\mp j/A$, and $\omega_j={e^{\pm 2\pi ij(j-k\,\mathrm{mod}\,A)/D}}$.
For later use it will also be useful to define $\tilde{B}_A^{(\Pi)}$ in a mixed position-momentum basis,
\begin{align}
\tilde{B}_{A,\mathrm{mix}}^{(\Pi)}:=F_D\tilde{B}_A^{(\Pi)}.
\end{align}
The associated classical map on the torus will be
\begin{align}\label{eqn:SA}
S_A^{(\Pi)}(x,p)&:= \Big(Ax-\lfloor Ax\rfloor, \frac{p+\Pi(\ell)}{A}\Big), \quad\text{for }x\in \Big[\frac{\ell}{A},\frac{\ell+1}{A}\Big),\; \ell=0,\ldots,A-1.
\end{align}
This is just the classical $A$-baker map with the permutation $\Pi$ applied to the stacking of the blocks shown on the right side of Figure~\ref{fig:s1}. Note that for $x\in [\frac{\ell}{A},\frac{\ell+1}{A})$, that $\ell=\lfloor Ax\rfloor$.

\subsection{Gaussian states on the torus}

We work on the $D$-dimensional Hilbert space of periodic states. Usually one would associate $F_{D/A}^{\alpha_j,\beta_j}$ with the Hilbert space of \emph{quasi}-periodic states corresponding to phase factors $\alpha,\beta$, but since we work with several different $\alpha_j,\beta_j$, there is no single particularly favored choice of phase conditions, and so we pick the periodic (no phase) boundary conditions for simplicity.
We first consider the $\R^2$-Gaussian state (also called ``coherent state'') localized at a point $\z_0=(x_0,p_0)\in\R^2$ with squeezing parameter $\sigma>0$,
\begin{align*}
\Psi_{\z_0,\sigma}(x)=(2D\sigma)^{1/4}e^{-\pi i Dx_0p_0}e^{2\pi i Dp_0x}e^{-\sigma D\pi(x-x_0)^2},\quad x\in\R.
\end{align*}
The corresponding torus coherent state $\Psi_{\z,\sigma,\T^2}\in\C^D$ is obtained by periodicizing the $\R^2$-state in position and momentum, resulting in
\begin{align}\label{eqn:torusst
ate}
\langle m|\Psi_{\z_0,\sigma,\T^2}\rangle &= \sum_{v\in\Z}\Psi_{\z_0,\sigma}\Big(\frac{m}{D}+v\Big),\quad m\in\Z/D\Z.
\end{align}
For further details, see \cite{bdb, DNW}.

The main properties we will use, in addition to the explicit formula for a Gaussian state, are the following.
\begin{enumerate}
\item For $0<\delta<\frac{1}{2}$ and $x_0\in(\delta,1-\delta)$, the torus coherent state $\Psi_{(x_0,p_0),\sigma,\T^2}$ is well-approximated by the $\R^2$-coherent state  $\Psi_{(x_0,p_0),\sigma}$: For $(\sigma D)\ge1$ and $j=0,\ldots,D-1$,
\begin{align}\label{eqn:R2approx}
\sqrt{D}\langle j|\Psi_{\z_0,\sigma,\T^2}\rangle &= \Psi_{\z_0,\sigma}\Big(\frac{j}{D}\Big)+\mathcal{O}\big((\sigma D)^{1/4}e^{-\pi\sigma D\delta^2}\big),
\end{align}
with a uniform implied constant in the error term.
This is due to the rapid decay of thet $\R^2$-Gaussian states, and is useful so we can apply the Gaussian transformation properties of the $\R^2$-states,
\begin{align*}
\psi_{(x_0,p_0),\sigma}(x+s)&=\psi_{(x_0-s,p_0),\sigma}(x)e^{\pi i Dp_0s}\\
\psi_{(x_0,p_0),\sigma}(rx)&=\psi_{(x_0/r,p_0r),r^2\sigma}(x)\frac{1}{\sqrt{r}}\\
\psi_{(x_0,p_0),\sigma}(x)&=\psi_{(x_0,p_0+g),\sigma}(x)e^{i\pi Dgx_0}e^{-2\pi iDgx}.
\end{align*}
\item The Fourier transform action on $\R^2$-Gaussian states translates to the torus Gaussian states through the DFT matrix,
\begin{align*}
F_D\Psi_{\z_0,\sigma,\T^2} &= \Psi_{F\z_0,1/\sigma,\T^2},
\end{align*}
where $F\z_0:=(p_0,-x_0)$. We will have to show the generalized DFT matrices $F_D^{\alpha,\beta}$ act in a similar enough way in the semiclassical limit.
For later use, we will use the following identity to relate $F_{D/A}^{\alpha,\beta}$ to $F_D$: For $j\in\{0,\ldots,A-1\}$ and $m-jD/A,\xi-\ell D/A\in\{0,\ldots,D/A-1\}$,
\begin{align}\label{eqn:FA}
\langle m-jD/A|F_{D/A}^{\alpha,\beta}|\xi-\ell D/A\rangle 
&=\sqrt{A}\langle Am-jD|F_D|\xi\rangle \, e^{-2\pi i\left(\beta\frac{A}{D}m+\alpha\frac{A}{D}\xi-j\beta-\alpha\ell+\alpha\beta\frac{A}{D}\right)}.
\end{align}

\item If $\z_0,\z_1\in\T^2$ are close enough, then the coherent states centered at $\z_0$ and $\z_1$ are close up to an overall phase factor:
If $\eta,\varepsilon=o(D^{-1/2})$ and $\sigma>0$ is fixed, then for $x_0\in(\delta,1-\delta)$ with $\delta\gg D^{-1/2}(\log D)^{-1/2}$, there is a phase factor $e^{i\Theta_{D,\eta,\varepsilon,\z_0}}$ so that
\begin{align}\label{eqn:cstate-bound}
\left\|\Psi_{(x_0,p_0),\sigma,\T^2}-e^{i\Theta_{D,\eta,\varepsilon,\z_0}}\Psi_{(x_0+\eta,p_0+\varepsilon),\sigma,\T^2}\right\|&=o(1),\quad\text{as }D\to\infty.
\end{align}
One can check this by estimating the euclidean norm directly, by going to the $\R^2$ Gaussian states and performing estimates depending on the index's distance from $Dx_0$. The requirements on $\eta,\varepsilon$ are intuitive, as the state $\Psi_{\z_0,\sigma,\T^2}$ is a Gaussian state with standard deviation of order $D^{-1/2}$ in position or momentum space, and so a state centered at $\z_0+o(D^{-1/2})$ eventually overlaps strongly with one at $\z_0$.
The requirement on $\delta$ is for convenience so we can interchange the torus states with the $\R^2$ Gaussian states according to~\eqref{eqn:R2approx}.

\hspace{3mm} This property is needed to handle the varying phases $\alpha_j,\beta_j$ in the DFT matrices; it is not required for the Balazs--Voros quantization with no phase factors.
\end{enumerate}

\subsection{Action on coherent states}

We consider $\z_0=(x_0,p_0)$ away from the discontinuities of the classical $A$-baker map; specifically for any $0<\gamma<\frac{1}{2}$ and $0<\delta<\frac{1}{2A}$, we consider
\[
\mathcal{G}_{A,\delta,\gamma}=\Big\{(x_0,p_0):\Big|x_0-\frac{k}{A}\Big|>\delta\;\;\forall k=0,\ldots,A,\text{ and }p_0\in(\gamma,1-\gamma)\Big\}.
\]
This partitions $[0,1]^2$ into $A$ disjoint rectangles, $I_\ell\times \Gamma:=\{x_0\in(\frac{\ell}{A}+\delta,\frac{\ell+1}{A}-\delta)\}\times\{p_0\in(\gamma,1-\gamma)\}$ for $\ell=0,\ldots,A-1$.
For each rectangle $I_\ell\times \Gamma$, we consider a coherent state centered at some $\z_0\in I_\ell\times \Gamma$, and compute its evolution under $\tilde{B}_A^{(\Pi)}$.

Suppose $\z_0\in I_\ell\times\Gamma$, and let $j=\Pi(\ell)$. Since the coherent state $\Psi_{\z_0,\sigma,\T^2}$ is localized at $\z_0$, it is small for coordinates away from $I_\ell$, in particular, using \eqref{eqn:R2approx}, then
\begin{align}\label{eqn:cstate-loc}
\langle \xi|\Psi_{\z_0,\sigma,\T^2}\rangle&=\frac{1}{\sqrt{D}}\mathcal{O}\big((\sigma D)^{1/4}e^{-\pi\sigma D\delta^2})\big),\quad \xi\not\in\left[\ell\frac{ D}{A},(\ell+1)\frac{D}{A}-1\right],
\end{align}
since $|\frac{\xi}{D}-x_0|\ge\delta$.
Thus the main contributions in $\tilde{B}_{A,\mathrm{mix}}^{(\Pi)}|\Psi_{\z_0,\sigma,\T^2}\rangle$ will come from the $\ell$th-block $\omega_\ell F_{D/A}^{\alpha_{\ell},\beta_{\ell}}$ in $\tilde{B}_{A,\mathrm{mix}}^{(\Pi)}$, since this block acts on the large elements of $|\Psi_{\z_0,\sigma,\T^2}\rangle$ which correspond to the indices $\xi\in[\ell\frac{D}{A},(\ell+1)\frac{D}{A}-1]$. This block has been permuted to the $j$th row block, and so we start with 
computing $\langle m|\tilde{B}_{A,\mathrm{mix}}^{(\Pi)}|\Psi_{\z_0,\sigma,\T^2}\rangle$ for $m\in[j\frac{D}{A},(j+1)\frac{D}{A}-1]$.
We are not concerned with phase factors that depend on $j$, $\ell$, $\alpha_{\ell}$, $\beta_{\ell}$, $D$, $A$, or $\z_0$ since these just become an overall phase factor, so we will accumulate them in a phase factor $e^{i\phi_{\Pi,\z_0}}$ which may change from line to line.
We will have to keep track of phase factors depending on $m$.
Using \eqref{eqn:FA}, \eqref{eqn:cstate-loc}, and Gaussian  transformation properties, we compute,
\begin{align}
\label{eqn:compute}
\begin{aligned}\langle m|\tilde{B}_{A,\mathrm{mix}}^{(\Pi)}|\Psi_{\z_0,\sigma,\T^2}\rangle
&=\omega_j\sum_{\xi=\ell\frac{D}{A}}^{(\ell+1)\frac{D}{A}-1}\langle m-jD/A|F_{D/A}^{\alpha_{\ell},\beta_{\ell}}|\xi-\ell D/A\rangle\langle \xi |\Psi_{\z_0,\sigma,\T^2}\rangle
\\
&=e^{i\phi_{\Pi,\z_0}}e^{-2\pi i \beta_{\ell}\frac{A}{D}m}\sqrt{A}\sum_{\xi=0}^{D-1}\langle Am-jD|F_D|\xi\rangle e^{-2\pi i\alpha_{\ell} \xi \frac{A}{D}}\langle \xi|\Psi_{\z_0,\sigma,\T^2}\rangle+\mathcal{O}(\sqrt{A}(\sigma D)^{1/4}e^{-\pi D\sigma\delta^2})\\
&=e^{i\phi_{\Pi,\z_0}}e^{-2\pi i \beta_{\ell}\frac{A}{D}m}\sqrt{A}\langle Am-jD|\Psi_{F(x_0,p_0-\alpha_{\ell}A/D),1/\sigma,\T^2}\rangle +\mathcal{O}(\sqrt{A}(\sigma D)^{1/4}e^{-\pi D\sigma\delta^2}),
\end{aligned}
\end{align}
where in the second line, we used \eqref{eqn:cstate-loc} to freely (at the cost of the error term $\mathcal{O}((\sigma D)^{1/4}e^{-\pi D\sigma\delta^2})$) add in the terms involving $\langle \xi|\Psi_{\z_0,\sigma,\T^2}\rangle$ for $\xi\not\in[\frac{\ell D}{A},\frac{(\ell+1)D}{A}-1]$, and in the third line we used
$e^{-2\pi i\alpha_{\ell}\xi\frac{A}{D}}\langle\xi|\Psi_{\z_0,\sigma,\T^2}\rangle =e^{-i\pi\alpha_{\ell}Ax_0}\langle \xi|\Psi_{(x_0,p_0-\alpha_{\ell}A/D),\sigma,\T^2}\rangle+\frac{1}{\sqrt{D}}\mathcal{O}((\sigma D)^{1/4}e^{-\pi\sigma D\delta^2})$ as well as the DFT action on the torus states.

By Gaussian state transformation properties, for $\z=(x_1,p_1)$,
\begin{align*}
\sqrt{A}\,\Psi_{F\z,1/\sigma}(Ax-j) &= \Psi_{((p_1+j)/A,-(Ax_1-\ell),A^2/\sigma}(x)e^{i\pi D \ell(p_1+j)/A}e^{-\pi i Djx_1}e^{-2\pi i D\ell x}.
\end{align*}
Then taking $x=m/D$ and $D$ large enough so $\gamma-A/D\ge \gamma/2$ (so that $p_0-\alpha_\ell A/D\in(\gamma/2,1-\gamma/2)$ and $\frac{p_0+j}{A}-\frac{\alpha_\ell}{D}\in(\frac{\gamma}{2A},1-\frac{\gamma}{2A})$), then
\begin{align*}
\sqrt{A}\langle Am-jD|\Psi_{F(x_0,p_0-\alpha_{\ell} A/D),1/\sigma,\T^2}\rangle= 
\langle m|\Psi_{(\frac{p_0+j}{A}-\frac{\alpha_{\ell}}{D},-(Ax_0-\ell)),A^2/\sigma,\T^2}\rangle e^{i\phi_{\Pi,\z_0}}+\mathcal{O}(\sigma^{-1/4}e^{-\pi D\gamma^2/(4\sigma)}).
\end{align*}
Using this in \eqref{eqn:compute} and taking into account the phase factor $e^{-2\pi i\frac{\beta_{\ell} A}{D}m}$, we obtain, for $m\in[j\frac{D}{A},(j+1)\frac{D}{A}-1]$,
\begin{align}\label{eqn:coord-est}
\langle m|\tilde{B}_{A,\mathrm{mix}}^{(\Pi)}|\Psi_{\z_0,\sigma,\T^2}\rangle &= e^{i\phi_{\Pi,\z_0}} \langle m|\Psi_{(\frac{p_0+j}{A}-\frac{\alpha_{\ell}}{D},-(Ax_0-\ell)-\frac{\beta_{\ell} A}{D}),A^2/\sigma,\T^2}\rangle+\mathcal{O}(\max(\sigma,1/\sigma)^{1/4}D^{1/2}e^{-\pi D\theta}),
\end{align}
where $\theta:=\max(\sigma\delta^2,\gamma^2/(4\sigma))$.

The same estimate holds for $m\not\in[j\frac{D}{A},(j+1)\frac{D}{A}-1]$, essentially because everything on the right side of \eqref{eqn:coord-est} is just $\mathcal{O}(\max(\sigma,1/\sigma)^{1/4}D^{1/2}e^{-\pi D\theta})$ as $|\frac{m}{D}-\frac{p_0+j}{A}|\ge\gamma/A$, and also so is the equivalent of the first line of \eqref{eqn:compute} by \eqref{eqn:cstate-loc}.
Then adding a factor $\sqrt{D}$ to go to the euclidean norm, we obtain,
\begin{align*}
\big\|\tilde{B}_{A,\mathrm{mix}}^{(\Pi)}\Psi_{(x_0,p_0),\sigma,\T^2}-e^{i\phi_{\Pi,\z_0}}\Psi_{(\frac{p_0+j}{A}-\frac{\alpha_{\ell}}{D},-(Ax_0-\ell)-\frac{\beta_{\ell} A}{D}),A^2/\sigma,\T^2}\big\| = \mathcal{O}\big(\max(\sigma,1/\sigma)^{1/4}D e^{-\pi D\theta}\big).
\end{align*}
Since the inverse DFT matrix $F_D^{-1}$ is unitary, then also
\begin{align}\label{eqn:normwitherror}
\big\|\tilde{B}_{A}^{(\Pi)}\Psi_{(x_0,p_0),\sigma,\T^2}-e^{i\phi_{\Pi,\z_0}}\Psi_{(Ax_0-\ell+\frac{\beta_{\ell}A}{D},\frac{p_0+j}{A}-\frac{\alpha_{\ell}}{D}),\sigma/A^2,\T^2}\big\| =\mathcal{O}\big(\max(\sigma,1/\sigma)^{1/4}D e^{-\pi D\theta}\big),
\end{align}
for $\theta=\max(\sigma\delta^2,\gamma^2/(4\sigma))$.
The above bound is $o(1)$ if for example, $\sigma\sim c$, $\gamma\sim c$, $\delta\sim\frac{c}{A}$, and $A=o(D^{1/2}(\log D)^{-1/2})$.
Finally, using \eqref{eqn:cstate-bound} with $\eta=\frac{\beta_{\ell} A}{D}$ and $\varepsilon=-\frac{\alpha_{\ell}}{D}$, yields,
\begin{align}\label{eqn:cstate-evolve}
\big\|\tilde{B}_{A}^{(\Pi)}\Psi_{(x_0,p_0),\sigma,\T^2}-e^{i\tilde{\phi}_{\Pi,\z_0}}\Psi_{(Ax_0-\ell,\frac{p_0+j}{A}),\sigma/A^2,\T^2}\big\|=o(1),\quad\text{as }D\to\infty.
\end{align}

\end{document}